\documentclass[%
  twoside,
  floatfix, 
  reprint,
  amsmath,amssymb,
  aps,
  pra,
  nofootinbib,
  superscriptaddress,
  a4paper
]{revtex4-2}

\usepackage{graphicx}
\usepackage[usenames,dvipsnames]{xcolor}
\usepackage{siunitx}
\usepackage{physics}
\usepackage{xfakebold} 

\usepackage{enumitem}
\setitemize{itemsep=0pt}

\usepackage[utf8]{inputenc}
\usepackage[T1]{fontenc}

\usepackage{lipsum}

\graphicspath{{Figures/}{}}

\definecolor{kclred}{RGB}{226,35,26}

\usepackage[
centering, includefoot,
text={7.1in,10.2in},
total={6.3in,8.75in},
top=0.8in, left=0.62in,
]{geometry}

\usepackage[
  bookmarks=true,
  colorlinks,
  linkcolor=blue,
  urlcolor=blue,
  citecolor=blue,
  plainpages=false,
  pdfpagelabels,
  final,
  breaklinks=true
]{hyperref}
\hypersetup{
pdftitle={Chiral moments make chiral measures}, 
pdfauthor={Emilio Pisanty, Nicola Mayer, Andrés Ordóñez, Alexander Löhr and Margarita Khokhlova}
}

\usepackage{natbib}

\AtBeginDocument{\RenewCommandCopy\qty\SI}
\ExplSyntaxOn 
\msg_redirect_name:nnn { siunitx } { physics-pkg } { none }
\ExplSyntaxOff

\renewcommand{\d}{\ensuremath{\textrm{d}}}
\renewcommand{\Re}{\operatorname{Re}}
\renewcommand{\Im}{\operatorname{Im}}
\DeclareSIUnit{\au}{{a.u.}}

\newcommand{\vbr}{\vb{r}}
\newcommand{\vbp}{\vb{p}}
\newcommand{\vbE}{\vb{E}}

\newcommand{\ue}[1]{\hat{\vb{e}}_{#1}}

\newcommand{\SO}[1]{\mathrm{SO}(#1)}

\makeatletter
\newcommand*{\tensordot}{}
\DeclareRobustCommand*{\tensordot}{%
  \mathbin{\mathpalette\tensordot@{}}%
}
\newcommand*{\tensordot@scalefactor}{.65} 
\newcommand*{\tensordot@widthfactor}{1.15}
\newcommand*{\tensordot@}[2]{%
  \sbox0{$#1\vcenter{}$}
  \sbox2{$#1\cdot\m@th$}%
  \hbox to \tensordot@widthfactor\wd2{%
    \hfil
    \raise\ht0\hbox{%
      \scalebox{\tensordot@scalefactor}{%
        \lower\ht0\hbox{$#1\bullet\m@th$}%
      }%
    }%
    \hfil
  }%
}
\makeatother

\newcommand{\tensorcross}{\setBold[0.7]\times\unsetBold}

\newcommand{\tens}[1]{\mathbf{#1}}
\newcommand{\id}{\mathbb{I}}
\NewDocumentCommand{\sym}{}{\trigbraces{\hat{\mathcal{S}}}}

\NewDocumentCommand{\lift}{}{\trigbraces{\hat{\mathcal{L}}}}

\newcommand{\proj}[1]{\trigbraces{\hat{\Pi}_{#1}}}
\newcommand{\projj}[2]{\trigbraces{\hat{\Pi}^{(#1)}_{#2}}}

\newcommand{\irrep}[2]{\mathcal{M}^{(#1)}_{#2}}
\newcommand{\symtens}[1]{\sym(\mathbb{C}^{d\otimes #1})}

\begin{document}

\title{Chiral moments make chiral measures}

\author{Emilio Pisanty}
\affiliation{Department of Physics, King’s College London, Strand Campus, London WC2R 2LS, UK}

\author{Nicola Mayer}
\affiliation{Department of Physics, King’s College London, Strand Campus, London WC2R 2LS, UK}

\author{Andrés Ordóñez}
\affiliation{Department of Physics, Freie Universit\"at Berlin, 14195 Berlin, Germany}

\author{Alexander Löhr}
\affiliation{Max Born Institute for Nonlinear Optics and Short Pulse Spectroscopy, Berlin, Germany}
\affiliation{Department of Physics, Humboldt University, Berlin, Germany}

\author{Margarita Khokhlova}
\affiliation{Department of Physics, King’s College London, Strand Campus, London WC2R 2LS, UK}

\date{\today}

\begin{abstract}
We develop a family of chiral measures to quantify the chirality of a distribution and assign it a handedness.
Our measures are built using the tensorial moments of the distribution, which naturally encode its spatial character, 
not only via its angular shape consistently with existing multipolar-moment approaches,
but also its radial dependence.
We combine these tensorial moments into a rotationally-invariant pseudoscalar using a newly-defined cross product and triple product for arbitrary symmetric tensors.
We analyze these measures for a variety of toy-model distributions, providing intuition for the geometry and guiding the choice of chiral measure optimal for a given distribution.
We also apply our measures to a physically-motivated example coming from photoionization in polychromatic chiral light.
Our work provides a robust, flexible, intuitive, highly geometrical, and physically-driven framework for understanding and quantifying the chirality of a wide variety of distributions,
together with an open-source software package that makes this toolbox readily applicable for the analysis of numerical or experimental data.
\end{abstract}
\maketitle

\section{Introduction}
Chirality is an essential concept in our description of the world, capturing physical shapes, structures and dynamics in three-dimensional space, by defining whether a given object is equivalent to its mirror image~\cite{Kelvin1894}.
Beyond abstract geometrical shapes,
we are surrounded by a kaleidoscope of chiral objects and structures:
from fundamental particles~\cite{Yang1958}
and nuclei~\cite{Ren2022}
all the way to cosmological structures~\cite{Cahn2023isotropic, Cahn2023test},
passing by
liquid crystals~\cite{Meyer1975, DeGennesProst1993},
nanoparticles and nanostructures~\cite{Hentschel2017, Gautier2009, Xie2025, Xie2025b, Vestler2018, Jones2023, Schulz2024, Mildner2023} 
and bulk solid structures~\cite{Fecher2022},
and, most importantly for human life, the homochiral biochemistry of organic molecules~\cite{Evans2012, Fujii2004, Blackmond2019}.
Even light itself can form chiral structures both in its spatial distribution~\cite{Tang2010, Bliokh2011} and in its evolution over time~\cite{Ayuso2019, Neufeld2020}, often with deep links to the foundations of electromagnetism~\cite{Cameron2017, Crimin2019}.

In all of these contexts, however, it is frequently not enough to know that an object is different from its mirror image: it is also necessary to understand \textit{how} different the two versions are~--
not just to identify chirality, but to quantify it.

A number of approaches have appeared over the years to fulfil this need \cite{Petitjean2003, Buda1992b, Fowler2005}.
Some rely on abstract geometry, quantifying the difference between a given shape and its mirror image~\cite{Gilat1989, Buda1992, Dryzun2011, Neufeld2020} 
or between the shape and an achiral reference~\cite{Buda1992b, Weinberg1993, Zabrodsky1995, Alvarez2005}.
In these geometrical approaches, a central challenge is the assignment of a handedness to a given object, i.e., actively distinguishing the two mirror versions from each other, and labelling them as `left-handed' or `right-handed'~\cite{Ruch1972}.

This problem is easier to solve in the chemical domain, 
either via conventional priority rules used to describe the structural backbone of a molecule~\cite{Cahn1956},
or through more observable-based approaches such as circular dichroism or optical rotatory power~\cite{Condon1937, Fernandez-Corbaton2016}.
However, those methods rarely extend to structures beyond the molecular realm,
setting the demand for a robust and universal measure of chirality,
which is applicable both to concrete shapes as well as to more general distributions, and thus directly relevant to the vast array of experiments where the measured observable is a distribution.

As a general rule, we demand from a measure of chirality that it 
(i) be independent of the orientation of the object,
(ii) be a continuous function of its shape, and
(iii) assign opposite signs to mirror-image objects.
For semantic clarity, for quantities that can detect chirality, but are unable to assign a handedness, we use the term `measure of asymmetry'.

An important and unavoidable feature of any such measures of chirality is the so-called `rubber-glove theorem'~\cite{Walba1991, Weinberg2011, Harris1999, Millar2005} (also known as `chiral connectedness'~\cite{Fowler2005, Mislow1993}):
informally, given a left-handed rubber glove, it is possible to pull it inside out, finger by finger, so that it becomes a right-handed glove without passing at any time through an achiral or mirror-symmetric configuration.
For a continuous chirality measure, this smooth passage between positive and negative values implies that at some point the measure must pass by zero, but, since the object is never achiral, that vanishing-measure point must occur for a chiral object:
in other words, the measure must have a `blind spot'.
The presence of such blind spots is generic and expected, with two important consequences.
Firstly, if we wish to quantify the chirality of a wide variety of shapes, we should expect to need a family of different measures~\cite{Rassat2004}, with different measures covering different blind spots.
Secondly, since different measures of the same family can disagree on the handedness of shapes, it is rarely possible to provide an absolute assignment of handedness, as it is challenging to give any priority within such a family.
That said, every member of the family still provides a valid sense of handedness for the shapes that it describes.

Perhaps the most attractive option currently available for quantifying the chirality of a distribution is based on the multipolar moments of the distribution~\cite{Harris1999, Osipov1995, Neal2003, Hattne2011}, which are then combined into a pseudo-scalar using the angular-momentum algebra of the rotation group, $\SO{3}$.
These provide robust and flexible measures, but their wide-scale application is held back by unintuitive forms and a lack of connection to the real-space geometries they represent, as well as to the physics embodied in those geometries.

In this work, we develop a robust family of chirality measures for arbitrary distributions, based on the distributions' tensorial moments $\mathbf{M}^{(k)}$.
We show that these measures take intuitive forms that are easily recognizable as `chiral moments',
given by newly-defined triple tensor products of the tensorial moments,
and taking the form
\begin{align}
\chi_{n_1 n_2 n_3}
=
\left( 
  \mathbf{M}^{(n_1)}
  \tensorcross
  \mathbf{M}^{(n_2)}
  \right)^{(n_3)}
\tensordot
\mathbf{M}^{(n_3)}
.
\nonumber
\end{align}
These factorized forms are simple enough to be calculated by hand, providing a clear link to the geometries and physical processes involved.
We show these measures in action on illustrative and didactic toy-model distributions.
Finally, as a concrete illustration, inspired by recent progress in ultrafast physics and increasing demand in chiral photoelectron spectroscopy~\cite{Mayer2022, Rajak2024, Comby2018, Fede2025, Katsoulis2022, Sparling2023, Sparling2024, Sparling2022, Sparling2025, Lux2012, Lux2015, Geyer2025, Powis1992, Nahon2015, Beaulieu2016, Han2025, Tikhonov2022, Heger2025, Kohnke2025}, we demonstrate their capabilities on the specific example of the photoelectron momentum distribution produced by resonance-enhanced multiphoton ionization of hydrogen driven by synthetic chiral light.

Our software implementation of this formalism is available as the open-source package \texttt{Chimera} in the Wolfram Language~\cite{pisanty-Chimera-2026}; 
the specific implementation of the results presented here is available at Ref.~\cite{figureMaker}.

\section{Chiral moments}
In one dimension, a distribution $\rho(x)$ is described by its moments, $M^{(n)}= \int x^n\rho(x)\d x$, a sequence of scalars measuring spatial aspects of its shape.
Similarly, to characterize a three-dimensional distribution $\rho(\vbr)$, one can use its tensorial moments,
\begin{align}
\mathbf{M}^{(n)}
=
\int
\vbr^{\otimes n}
\rho(\vbr)
\d\vbr
,
\label{M-tensor-definition}
\end{align}
where $\vbr^{\otimes n} = \vbr \otimes \cdots \otimes \vbr$ is the $n$-fold tensor product of $\vbr$ with itself, and $n$ is the rank of the tensorial moment.
This tensor of rank $n$ is the object whose components in a Cartesian frame are
\begin{align}
M^{(n)}_{i_1\cdots i_n}
=
\int
x_{i_1} \cdots x_{i_n}
\rho(\vbr)
\d\vbr
,
\end{align}
with each of the $n$ indices, $i_1$ through $i_n$, ranging independently over $x,y,z$.
We include, in Appendix~\ref{sec-appendix-tensor-technical}, a brief summary of the relevant tensor algebra.

The so-called `unabridged' or `primitive' tensorial moments $\mathbf{M}^{(n)}$~\cite{Raab2005Multipole, Applequist1984, Zangwill2013} are typically reduced to the tensorial multipolar moments~\cite{Raab2005Multipole, Applequist1984}
\begin{align}
\boldsymbol{\mu}^{(\ell)} = \hat{\Pi}_\ell \mathbf{M}^{(\ell)}
,
\label{mu-ell-definition}
\end{align}
using a trace-removal projector $\hat{\Pi}_\ell$, 
which we motivate and define in Appendix~\ref{sec-appendix-tensor-technical};
the tensorial moment of rank $n=\ell$ provides the $2^\ell$-polar moment.
The tensorial multipolar moments form irreducible representations of the rotation group $\SO{3}$, and they are fully determined by a minimal set of linearly-independent components,
as we explore in detail in Appendix~\ref{app-multipolar-tensor-technical}.
These components are the spherical multipolar moments
\begin{align}
M_{\ell m}
& =
\int
S_{\ell m}(\vbr)
\rho(\vbr)
\d\vbr
,
\label{rho-lm-definition}
\end{align}
defined as integrals against the solid harmonic polynomial
$
S_{\ell m}(\vbr)
=
\sqrt{
  \frac{4\pi}{2\ell+1}
  }
\:
r^\ell
\:
Y_{\ell m}(\theta,\phi)
$.

As a general rule, the spherical multipolar moments $M_{\ell m}$ capture the details of the angular shape of the distribution $\rho(\vbr)$ projected to the unit sphere.
The unabridged moment tensors $\mathbf{M}^{(\ell)}$, on the other hand, also capture this information, while including additional information about the radial structure of the distribution.
Due to their simplicity and generality, we aim to build our chiral measure directly from the latter.

The chiral measure has to be a single pseudoscalar, which we build out of the tensorial moments, as a linear combination of products of their components.
We find that the simplest such pseudoscalar requires three tensorial moments, in the shape of a tensorial triple product, 
which we define for symmetric tensors as
\begin{align}
\left( 
  \mathbf{A}^{(n_1)}
  \tensorcross
  \mathbf{B}^{(n_2)}
  \right)^{(n_3)}
\tensordot
\mathbf{C}^{(n_3)}
& =
\epsilon_{i_1 i_2 i_3}
A_{i_1 \mathbf{j} \mathbf{k}}
B_{i_2 \mathbf{j} \mathbf{l}}
C_{i_3 \mathbf{k} \mathbf{l}}
,
\label{tensor-triple-product-definition}
\end{align}
as we detail in Appendices~\ref{app-definitions-and-properties} and \ref{sec-appendix-tensor-technical}.
Here
$\mathbf j = (j_1\cdots j_{m})$,
$\mathbf k = (k_1\cdots k_{n})$ and
$\mathbf l = (l_1\cdots l_{p})$
are multi-indices of lengths
$m=\frac12(n_1 + n_2 - n_3 -1)$,
$n=\frac12(n_1 + n_3 - n_2 -1)$ and 
$p=\frac12(n_2 + n_3 - n_1 -1)$.
This tensorial triple product has the form of a full contraction of the third factor, $\mathbf{C}^{(n_3)}$, with a newly-defined tensorial cross product 
  $
\left( 
  \mathbf{A}^{(n_1)}
  \tensorcross
  \mathbf{B}^{(n_2)}
  \right)^{(n_3)}_{i_3 \mathbf{k} \mathbf{l}}
=
\sym_{i_3 \mathbf{k} \mathbf{l}}
\epsilon_{i_1 i_2 i_3}
A_{i_1 \mathbf{j} \mathbf{k}}
B_{i_2 \mathbf{j} \mathbf{l}}
$,
where $\sym_{i_3 \mathbf{k} \mathbf{l}}$ denotes full symmetrisation over the indices $i_3 \mathbf{k} \mathbf{l}$,
which we define in detail in Appendix~\ref{sec-appendix-tensor-technical}.
We thus have a family of chiral measures, 
\begin{align}
\chi_{n_1 n_2 n_3}
=
\left( 
  \mathbf{M}^{(n_1)}
  \tensorcross
  \mathbf{M}^{(n_2)}
  \right)^{(n_3)}
\tensordot
\mathbf{M}^{(n_3)}
,
\label{chi-initial-definition}
\end{align}
coming from the different possible combinations of tensor ranks $n_i$, and which carry the imprint of different types of chiral shapes of the distribution.

These chiral measures can be re-expressed into more straightforward forms by expanding the integral form~\eqref{M-tensor-definition} of the individual tensor factors, which gives us the equivalent expressions
\begin{subequations}%
\begin{align}
\chi_{n_1 n_2 n_3}
& =
\iiint
\d\vbr_1
\d\vbr_2
\d\vbr_3
\rho(\vbr_1)
\rho(\vbr_2)
\rho(\vbr_3)
\\ & \qquad \qquad \quad \nonumber
\left( 
  \vbr_1^{\otimes n_1}
  \tensorcross
  \vbr_2^{\otimes n_2}
  \right)^{(n_3)}
\tensordot
\vbr_3^{\otimes n_3}
\\
& =
\iiint
\d\vbr_1
\d\vbr_2
\d\vbr_3
\rho(\vbr_1)
\rho(\vbr_2)
\rho(\vbr_3)
\\ & \qquad \quad \nonumber
\left[
  (\vbr_1 \times \vbr_2) \cdot \vbr_3 
  \right]
(\vbr_1 \cdot \vbr_2)^{m}
(\vbr_1 \cdot \vbr_3)^{n}
(\vbr_2 \cdot \vbr_3)^{p}
.
\end{align}
\label{chiral-moment-as-triple-integral}%
\end{subequations}%
Our chiral measures are thus revealed as
three-point correlation functions (three copies of $\rho(\vbr)$ evaluated at different points, integrated against a chiral correlation kernel),
and they have the general structure of a moment: a distribution integrated against a polynomial.
We therefore call our chiral measure $\chi_{n_1 n_2 n_3}$ a `chiral moment' of the distribution~$\rho(\vbr)$.

A similar chiral moment can be formed from the traceless version of the tensorial moments, the tensorial multipolar moments,
given naturally as
\begin{align}
h_{\ell_1 \ell_2 \ell_3}
=
\left( 
  \boldsymbol{\mu}^{(\ell_1)}
  \tensorcross
  \boldsymbol{\mu}^{(\ell_2)}
  \right)^{(\ell_3)}
\tensordot
\boldsymbol{\mu}^{(\ell_3)}
.
\label{h-l1l2l3-definition}
\end{align}
This traceless chiral moment $h_{\ell_1 \ell_2 \ell_3}$ is a pseudoscalar formed from three $\SO{3}$ irreducible representations.
As such, the representation theory of $\SO{3}$ constrains it~\cite{Hall2004} to have the form
\begin{equation}
h_{\ell_1 \ell_2 \ell_3}
\propto
\sum_{m_1 m_2 m_3}
\!\!\!
\begin{pmatrix}
  \ell_1 & \ell_2 & \ell_3 \\
  m_1 & m_2 & m_3
\end{pmatrix}
M_{\ell_1 m_1} 
M_{\ell_2 m_2}
M_{\ell_3 m_3}
,
\end{equation}
as a linear combination of the linearly-independent components, the spherical multipolar moments $M_{\ell m}$,
with linear-combination coefficients given by Wigner $3j$ symbols.
This form is the sharpest existing expression of the formalism in the literature, and it has been used previously from molecular chirality~\cite{Harris1999, Osipov1995, Neal2003, Hattne2011, Papakostas2003, Potts2003} through to cosmological contexts~\cite{Cahn2023isotropic, Cahn2023test}.
For the traceless chiral moments, the correlation kernel takes the form
$
\left( 
  \hat{\Pi}_{\ell_1}
  \vbr_1^{\otimes \ell_1}
  \tensorcross
  \hat{\Pi}_{\ell_2}
  \vbr_2^{\otimes \ell_2}
  \right)^{(\ell_3)}
\tensordot
\hat{\Pi}_{\ell_3}
\vbr_3^{\otimes \ell_3}
$,
and it can be shown to equal the scalar tripolar spherical harmonic~\cite{Varshalovich1988}; 
we provide explicit forms in Appendix~\ref{app-explicit-forms}.

On the intuitive side, the tensorial triple product behaves similarly to the usual triple product between vectors, $(\vb{a}\times\vb{b})\cdot\vb{c}$.
Thus, in the same way that the cross product of a vector with itself vanishes,
$
\vb{a}\times \vb{a} = 0
,
$ 
the cross product of a tensor with itself also vanishes,
$
\left( 
  \mathbf{M}^{(n)}
  \tensorcross
  \mathbf{M}^{(n)}
  \right)^{(n')}
  =
  0
,
$ 
regardless of the desired rank $n'$.
As a consequence, obtaining a nonzero tensorial triple product (and, therefore, a nonzero chiral moment) requires taking three different tensorial factors.

To satisfy this, there are two distinct options.
The first option is to assemble the chiral moment out of three different ranks of tensorial moment, in which case the first nontrivial example is $\chi_{234}$.%
\footnote{
The next nontrivial examples are $\chi_{245}$, $\chi_{256}$ and $\chi_{346}$.
}
The second option is to assemble the chiral moment using tensor factors of equal ranks, e.g.\ attempting to build $\chi_{111}$, 
in which case we must use tensorial moments with additional radial factors,
introducing powers of $r^2$ into the definition of $\mathbf{M}^{(n)}$ to yield tensor-valued moments of higher polynomial orders,
\begin{align}
\mathbf{M}^{(n,2q)}
=
\int
\vbr^{\otimes n}
r^{2q}
\rho(\vbr)
\d\vbr
=
\Tr^q(\mathbf{M}^{(n+2q)})
,
\label{M-n-2q-definition}
\end{align}
which are directly given as the $q$-fold trace of the higher-order tensorial moment $\mathbf{M}^{(n+2q)}$.
By extension, these also give rise to higher-order traceless tensorial multipolar moments
$
\boldsymbol{\mu}^{(\ell,2q)} = \hat{\Pi}_\ell \mathbf{M}^{(\ell,2q)}
$%
.%
\footnote{
As an alternative approach to achieve this, one can restrict the integration to different spherical shells for the different tensorial-moment factors in $\chi_{n_1n_2n_3}$;
this yields useful chiral moments which will be explored in future work.
}
In what follows, we explore both options.

\section{Analytical examples}
We now turn to specific examples of chiral distributions, to show our chiral moments in action as chiral measures.
We start with the simplest example of a chiral distribution, and we work our way up in complexity.

We construct our examples using gaussian distributions. For notational convenience, we write an arbitrary gaussian distribution as
\begin{align}
G_\Sigma(\vbr)
=
{
  (2\pi \det(\Sigma))^{-3/2}
  }
\ 
e^{-\frac12\vbr\Sigma^{-1}\vbr}
,
\end{align}
centred at the origin, where $\Sigma_{ij}$ is the covariance matrix of the distribution, which determines its width and orientation, and which can be shown to equal the second-order moments
\begin{align}
\Sigma_{ij}
=
\int
x_ix_j
G_\Sigma(\vbr)
\d\vbr
.
\end{align}
When the reference frame is aligned with the principal axes of the gaussian, this takes the form $\Sigma = \mathrm{diag}(\sigma_1^2, \sigma_2^2, \sigma_3^2)$, in which case its diagonal entries, the variances $\sigma_i^2$, are the squares of the corresponding standard deviations $\sigma_i$.

\subsection{The triple-dipole case: \texorpdfstring{$\chi_{111}$}{χ(111)}}
\label{sec-chi-111}
\begin{figure}[t]
    \centering
    \includegraphics[width=0.7\linewidth]{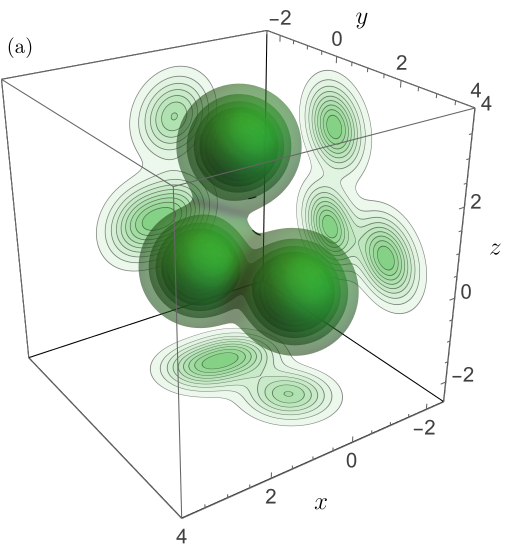}
    \includegraphics[width=0.7\linewidth]{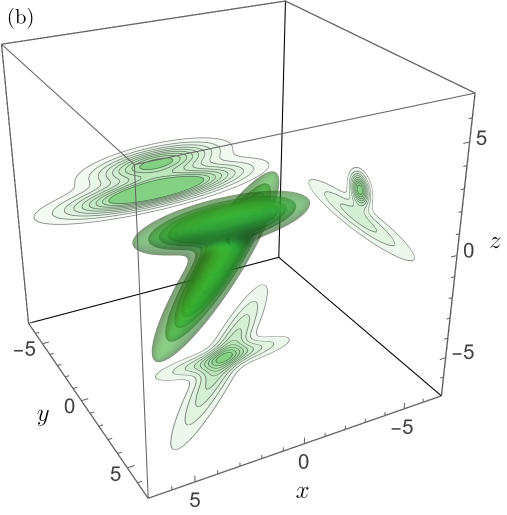}
    \caption{
    (a) Triple-gaussian distribution $\rho_\mathrm{3G}(\vbr)$, with the gaussians centred at $\vbr_1 = \ue{x}$, $\vbr_2 = 2\ue{y}$ and $\vbr_3 = 3\ue{z}$, and with standard deviation $\sigma=1$.
    The central 3D figure shows contours of equal probability, with the auxiliary projections to each plane showing the corresponding marginal distribution.
    (b) \textit{Vellela vellela}-type distribution $\rho_{vv}(\vbr)$, with two vertically-displaced elongated gaussians, setting $\sigma_1=2$, $\sigma_2=4$, $\sigma_\mathrm{sm}=1/2$ and $z_0 = -3/2$, with twist angle $\theta = \SI{45}{\degree}$.
    }
    \label{fig:distributions-111-and-221}
\end{figure}

The simplest chiral distribution is a combination of three spherical gaussians of equal variance $\sigma^2$ at different positions,
\begin{align}
\rho_\mathrm{3G}(\vbr)
&=
\frac13
\sum_{i=1}^3
G_{\sigma^2\mathbb{I}}(\vbr-\vbr_i)
,
\end{align}
as shown in Fig.~\ref{fig:distributions-111-and-221}(a).
Intuitively, so long as the vectors from the origin to the three centres of the gaussians form a non-coplanar set, with a nonzero triple product, 
$(\vbr_1\times\vbr_2)\cdot\vbr_3$,
the distribution itself is chiral.

It is tempting to attempt to extract this triple product of the centres through the triple tensorial product of rank $n=1$, $(\mathbf{M}^{(1)} \times \mathbf{M}^{(1)})^{(1)} \tensordot \mathbf{M}^{(1)}$, but, as discussed above, this vanishes due to its symmetry.
To build a suitable chiral measure, then, we use the tensorial moments with added radial factors $\mathbf{M}^{(n,2q)}$ as defined in~\eqref{M-n-2q-definition}.%
\footnote{
In nuclear physics, the moment $\tens{M}^{(1,2)} = \int r^2 \:\vbr \: \rho(\vbr)\d\vbr$ is known as the Schiff moment~\cite{Schiff1963, Flambaum2020}.
}
In this spirit, then, we define
\begin{align}
\chi_{111}^{024}
=
\left( 
  \mathbf{M}^{(1)}
  \tensorcross
  \mathbf{M}^{(1,2)}
  \right)^{(1)}
\tensordot
\mathbf{M}^{(1,4)}
.
\end{align}
For our triple-gaussian example, it is a simple calculation to show that
%
%
\begin{align}
\chi_{111}^{024}
&=
\frac{1}{27}
(r_1^2 - r_2^2)
(r_2^2 - r_3^2)
(r_3^2 - r_1^2)
(\vbr_1\times\vbr_2)\cdot\vbr_3
.
\end{align}
As expected, this returns a vanishing chirality if any of the radii coincide.%
\footnote{
Moreover, the scalar factor 
$
(r_1^2 - r_2^2)
(r_2^2 - r_3^2)
(r_3^2 - r_1^2)
$
can be recognised as the Vandermonde determinant 
$
\det\mathopen{}\begin{pmatrix}
1 & 1 & 1 \\
r_1^2 & r_2^2 & r_3^2\\
r_1^4 & r_2^4 & r_3^4
\end{pmatrix}
$.
}

For this triple-gaussian case, it is important to note that the chiral measure $\chi_{111}^{024}$ only makes sense for distributions for which the origin of the variable $\vbr$ is well defined.
This is the case e.g.\ for photoelectron momentum spectra, where $\vbp=0$ corresponds to zero kinetic energy.
However, there are also many cases of distributions, particularly spatial ones, where translational invariance makes this measure meaningless.%
\footnote{%
A related chirality measure is available that avoids this issue. 
For a chiral configuration of $N \geq 4$ gaussians, the origin can be set at the centre of mass, such that $\tens{M}^{(1)}=0$.
In this case, $\chi_{111}^{246}$ is an appropriate chirality measure and is now independent of the initial location of the origin.
}

As a brief note, in this case, the corresponding traceless chiral moment $h_{111}^{024}$ coincides with $\chi_{111}^{024}$.

\subsection{The double-quadrupole case: \texorpdfstring{$\chi_{221}$}{χ 221}}
\label{sec-chi-221}
We now climb one step up in complexity, to a distribution with quadrupolar ($\ell=2$) structure.
The first nontrivial chirality measure here is $\chi_{221}$, which, as before, contains two repeated $\ell$'s, i.e., two independent quadrupole moments.
The double-quadrupole chiral measure requires adding a radial factor to one of the tensorial moments:
\begin{align}
\chi_{221}^{020}
=
\left( 
  \mathbf{M}^{(2)}
  \tensorcross
  \mathbf{M}^{(2,2)}
  \right)^{(1)}
\tensordot
\mathbf{M}^{(1)}
.
\end{align}

To construct an explicit example of a distribution with nonzero $\chi_{221}^{020}$, we take two elongated gaussians, with different lengths (to make $\mathbf{M}^{(2)}$ and $\mathbf{M}^{(2,2)}$ distinct), at an angle $\theta$ to each other, and we displace them along the direction of the tensor cross product
$
\left( 
  \mathbf{M}^{(2)}
  \tensorcross
  \mathbf{M}^{(2,2)}
  \right)^{(1)}
$,
as shown in Figure~\ref{fig:distributions-111-and-221}.
Thus, we take
\begin{align}
\rho_{vv}(\vbr)
& =
w_1
G_{\Sigma_1}(\vbr)
+
w_2
G_{R_z(\theta)\Sigma_2R_z^{-1}(\theta)}(\vbr-\vbr_0)
,
\end{align}
where the covariance matrices of the two components are given by 
$\Sigma_i = \mathrm{diag}(\sigma_i^2, \sigma_\mathrm{sm}^2, \sigma_\mathrm{sm}^2)$,
with a common small-axis variance $\sigma_\mathrm{sm}^2$,
$\Sigma_2$ has been rotated by an angle $\theta$ about the $z$ axis through the matrix $R_{z}\mathopen{}(\theta)\mathclose{}$,
the second gaussian has been displaced by 
$\vbr_0=z_0\ue{z}$, 
and the weights $w_i$ add to $w_1+w_2=1$.

We encounter here our first cross product between tensors,%
\footnote{%
This cross product between rank-2 tensors has been used previously in continuum mechanics~\cite{Szabo2016, Bonet2015, Altenbach2012}, introduced (to our knowledge) in Ref.~\cite{deBoer1982}.
}
and, for this case, it essentially captures the cross product between the major axes of the two gaussians,
while allowing for their ambiguity in sign,
and it is given~by
\begin{align}
\left(
  \mathbf{M}^{(2)}
  \tensorcross
  \mathbf{M}^{(2,2)}
  \right)^{(1)}
& =
C
w_1 w_2
\sin(2\theta)
\ue{z}
,
\end{align}
where
$
C
=
\frac12
(z_0^2-3\sigma_1^2+3\sigma_2^2) 
(\sigma_1^2-\sigma_\mathrm{sm}^2) 
(\sigma_2^2-\sigma_\mathrm{sm}^2)  
$.
When contracted with the dipole moment of the distribution,
$
\mathbf{M}^{(1)}
=
w_2z_0\ue{z}
$,
this gives us the chiral moment
\begin{align}
\chi_{221}^{020}
=
C
w_1 w_2^2
z_0
\sin(2\theta)
.
\end{align}
The dependence on $\sin(2\theta)$ captures the fact that at $\theta=\SI{90}{\degree}$ the distribution admits mirror symmetry planes along the $xz$ and $yz$ axes, and is therefore achiral.

We get an identical value for the traceless chiral moment $h_{221}^{020}$, since the traceless quadrupole moments $\boldsymbol{\mu}^{(2)}$ differ from the $\mathbf{M}^{(2)}$ by a factor of the isotropic tensor $\mathbb{I}$, whose cross products vanish.

This type of chirality is exhibited in nature by the hydrozoan \textit{Velella velella}~\cite{Cameron2023, Cameron2024} (also known as by-the-wind sailor),
and it is also seen in man-made objects such as oblique-wing aircraft~\cite{Mustard2024};
it also corresponds to molecules with two independent quadrupole tensors, such as polarizabilities at different frequencies, and a permanent dipole moment~\cite{Cameron2023, Cameron2024}.

\subsection{The helical case: \texorpdfstring{$\chi_{234}$}{χ 234}}
\label{sec-chi-234}
As our third analytical example, we now turn to the first chiral measure which is fully flexible, $\chi_{234}$. 
The flexibility comes from involving three different tensor ranks, which allows measuring the chirality of a distribution confined to a single spherical shell.
That stands in contrast to our previous examples, which required including the radial structure of the distribution (by replacing $\mathbf{M}^{(n)}$ with $\mathbf{M}^{(n,2q)}$) and thus can only describe chirality which is spread over multiple spherical `shells' of different radii.
In essence, this measure is formed from the combination of a quadrupole, octupole and hexadecapole moments~-- which can also be taken as traceless moments via $h_{234}$~\cite{Harris1999}~-- and captures their relative shape and orientation.

In geometrical terms, this chirality measure is best suited to describe helical structures, and carries the imprint of the local pitch of the helix, which can be double- or triple-stranded.
That said, given the wider variety of geometries captured by octupolar and hexadecapolar moments, this chiral moment also extends to a wider class of shapes.

\begin{figure}[t]
    \centering
    \includegraphics[width=0.7\linewidth]{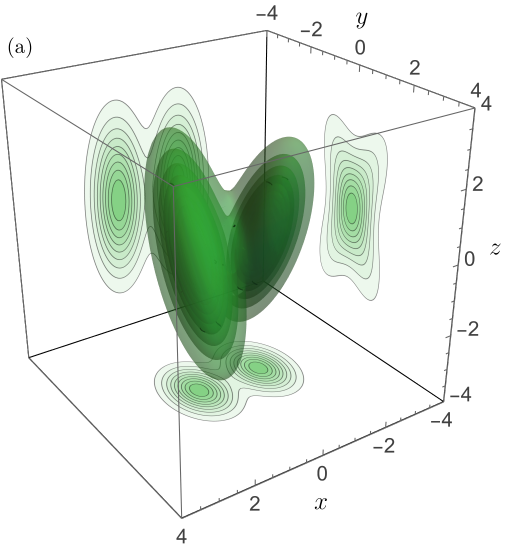}
    \includegraphics[width=0.7\linewidth]{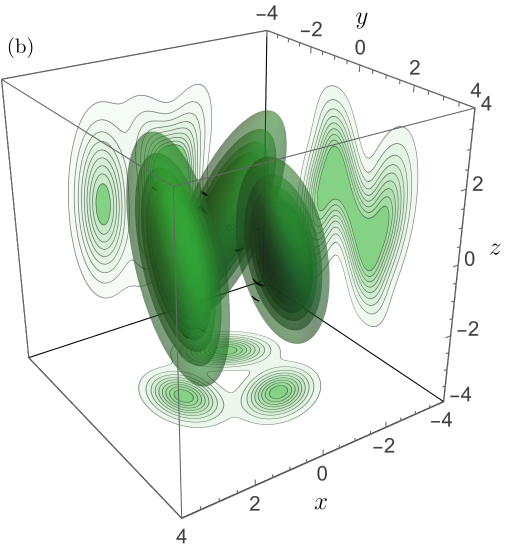}
    \caption{
    Helix-gaussian distributions $\rho_{\mathrm{helix}}(\vbr)$ for (a) $N=2$ and (b) $N=3$ gaussians.
    We set $\sigma_1=1/2$, $\sigma_2=3/2$, $r_0=1$, and use twist angle $\alpha = \SI{22.5}{\degree}$.
    }
    \label{fig:distributions-helix}
\end{figure}

To provide an explicit analytical example $\rho_{\mathrm{helix}}(\vbr)$, we consider the combination of $N=2,3$ gaussian distributions centred at equispaced points in the $(x,y)$ plane at radius $r_0$, with variances $(\sigma_1^2, \sigma_1^2, \sigma_2^2)$, and rotated by an angle $\alpha$ about the axis that connects them to the origin.
This distribution thus has the probability density 
\begin{align}
\rho_{\mathrm{helix}}(\vbr)
&=
\frac{1}{N}
\sum_{j=1}^N
G_{R_{x}(\alpha)\Sigma^{-1}R_{x}^{-1}(\alpha)}\mathopen{}
\left(
  R_{z}^{-1}(\tfrac{2\pi j}{N})
  \vbr
  -\vbr_0
  \right)
,
\end{align}
where the covariance matrix 
$\Sigma = \mathrm{diag}(\sigma_1^2, \sigma_1^2, \sigma_2^2)$
is rotated along the $x$ axis by $R_{x}\mathopen{}(\alpha)\mathclose{}$,
and the centroid $\vbr_0=r_0\ue{x}$ 
is then rotated about the $z$ axis by $R_{z}\mathopen{}(\tfrac{2\pi j}{N})\mathclose{}$.
We show examples in Figure~\ref{fig:distributions-helix}.

For the case of $N=2$ gaussians, the chiral moment can be calculated analytically to be
\begin{align}
\chi_{234}
&=
-
\frac{1}{16}
r_{0}
\Delta^2
\bigg[
  8 r_0^2(r_0^2+ \Delta)
    \\ \nonumber & \qquad \qquad \qquad 
  +3 \Delta^2 (1- \cos(4 \alpha))
  \bigg]
\sin(4 \alpha)
,
\end{align}
where $\Delta=\sigma_1^2-\sigma_2^2$,
with the traceless chiral moment given by
\begin{align}
h_{234}
&=
-
\frac{1}{112}
r_{0}
\Delta^2
\bigg[
  8 r_0^2(7r_0^2+9\Delta)
    \\ \nonumber & \qquad \qquad \qquad 
  +21 \Delta^2 (1- \cos(4 \alpha))
  \bigg]
\sin(4 \alpha)
.
\end{align}
As in the previous case, the distribution is achiral when $\alpha=\SI{45}{\degree}$.

Similarly, for the case of $N=3$ gaussians, the chiral moment is given by
\begin{align}
\chi_{234}
& =
-\frac{3}{128} 
r_0
\Delta
\sin(2\alpha) 
\left(
  2r_0^2
  +\Delta(1{-}\cos(2\alpha))
  \right)
 \\ &  \times \nonumber
\left(
  4r_0^2
  (r_0^2{+}2\Delta)
  +
  3
  \Delta^2
  (1{+}3\cos(2\alpha))
  (1{-}\cos(2\alpha))
  \right)
.
\end{align}
In contrast to the $N=2$ case, however, for the $N=3$ geometry the traceless chiral moment coincides with the unabridged one, i.e., $h_{234}=\chi_{234}$.

This difference in behaviour between the $N=2$ and $N=3$ geometries is remarkable and merits a closer look.
The two chiral moments involved, 
$\chi_{234}$ (in terms of $\mathbf{M}^{(n)}$
and
$h_{234}$ (in terms of $\boldsymbol{\mu}^{(\ell)}$)
contain largely similar ingredients.
The key difference between them is the passage from $\mathbf{M}^{(4)}$ to $\boldsymbol{\mu}^{(4)}$, which differ from each other by a term proportional to $\sym(\boldsymbol{\mu}^{(2,2)} \otimes \id)$ 
(see Appendix~\ref{sec-appendix-tensor-technical} for details)%
.
This term then shows up in the difference between the two chiral moments,%
\footnote{
Specifically, 
$
\chi_{234} - h_{234} 
=
\left( 
  \boldsymbol{\mu}^{(2)}
  \tensorcross
  \boldsymbol{\mu}^{(3)}
  \right)^{(4)}
\tensordot
\tfrac{6}{7}
\sym(\boldsymbol{\mu}^{(2,2)} \otimes \id)
$,
which can then be rearranged into 
$
-\tfrac{2}{7}
\left( 
  \boldsymbol{\mu}^{(2)}
  \tensorcross
  \boldsymbol{\mu}^{(2,2)}
  \right)^{(3)}
\tensordot
\boldsymbol{\mu}^{(3)}
$.
}
giving rise to a lower-rank chiral moment:
\begin{align}
\chi_{234} - h_{234} 
& =
-\tfrac{2}{7}
h_{223}^{020}
=
-\tfrac{2}{7}
\chi_{223}^{020}
,
\label{chi234-minus-h234-to-chi234020}
\end{align}
independently of $N$.

This last chiral moment is an interesting and significant chiral measure in its own right, and it provides a particularly simple chiral measure for $\rho_\mathrm{helix}(\vbr)$ in the case of $N=2$, where it is given by
\begin{align}
\chi_{223}^{020}
& =
-
\frac12
r_0^3
\Delta^3
\sin(4 \alpha)
.
\end{align}
This case represents the simplest instance where a rank-3-valued tensor cross product relates directly to critical physical and geometrical features, 
more specifically the tensor cross product
\begin{align}
\left(
  \mathbf{M}^{(2)}
  \tensorcross
  \mathbf{M}^{(2,2)}
  \right)^{(3)}
& =
-
2
r_0^2
\Delta^2
\cos(2\alpha)
\:
\sym(
  \ue{x} \otimes \ue{y} \otimes \ue{z}
  )
,
\end{align}
valid for $N=2$.
This is a tensor cross product between rank-2 tensors, both of which are diagonal in the $xyz$ reference frame, but which have different eigenvalues along those principal axes.

By contrast, for the $N=3$ geometry, the two quadrupole tensors $\mathbf{M}^{(2)}$ and $\mathbf{M}^{(2,2)}$ are linearly dependent, 
because the three-fold symmetry of $\rho_\mathrm{helix}(\vbr)$ requires both tensors to be axially symmetric about the $z$ axis.
Since the two are linearly dependent, their cross product vanishes, as (therefore) does the difference between $\chi_{234}$ and $h_{234}$.

\section{Chiral photoelectron momentum distributions}
\label{sec-photoelectron-spectra}
Having explored the application of our chiral measures to toy-model distributions, we now turn to a concrete physical object.
Our object of choice is the photoelectron momentum distribution arising from ionization by synthetic chiral light~\cite{Ayuso2019, Mayer2022, Katsoulis2022, Geyer2025}.

Light has been a useful tool to probe chiral matter since the introduction of chirality~\cite{Pasteur1905, Kelvin1894}.
Historically, light used to probe chiral matter has been circularly polarized, for which the chiral optical interaction is based on magnetic-dipole and electric-quadrupole effects, where the optical chirality is described through the spatial pitch of the circular-polarization helix.
Methods based on this interaction, including optical rotation and circular dichroism \cite{Condon1937}, are effective and form the gold standard for optically thick samples.

For optically thin samples, on the other hand, the length-scale mismatch between the wavelength-scale helical pitch and the size of typical chiral molecules makes these effects relatively weak~\cite{Ayuso2022road}.
For these cases, chiral photoelectron spectroscopy has emerged over the past two decades, allowing the use of chiral experimental configurations (such as the detection of forward-backward asymmetry in photoelectron emission~\cite{Mayer2022, Powis2008, Rajak2024, Comby2018, Fede2025, Katsoulis2022, Sparling2023, Sparling2024, Sparling2022, Sparling2025, Lux2012, Lux2015, Geyer2025, Powis1992, Nahon2015, Beaulieu2016, Heger2025}) to power all-electric-dipole methods that are highly enantiosensitive even at single-molecule scale, and opening access to time-resolved studies of chiral dynamics~\cite{Tikhonov2022, Ayuso2022road}.

Within this vein, recent work has exhibited both photoelectron circular dichroism~\cite{Sparling2025} as well as richly-structured three-dimensional chiral photoelectron momentum distributions, for 
single-photon ionization~\cite{Tia2017, Fehre2021}, 
resonantly-enhanced multiphoton ionization~\cite{Lux2012, Comby2018} (REMPI), as well as 
strong-field ionization~\cite{Rajak2024, Bloch2021, Mayer2022, Fede2025, Ordonez2025}.
Moreover, these methods offer additional promise when driven by synthetic chiral light~\cite{Ayuso2019, Habibovic2024, Kohnke2025}, a three-dimensional polychromatic combination with chiral sub-cycle dynamics.
These emerging tools, and the investigation of a plethora of new processes enabled by them, 
generally encode rich structural and dynamical information into the three-dimensional structure of the photoelectron momentum distribution, 
thus creating the need for tools to characterize the chiral features of the latter.

\begin{figure*}[t!]
    \begin{tabular}{cc}
      \raisebox{0mm}{%
        \includegraphics[height=0.28\linewidth]{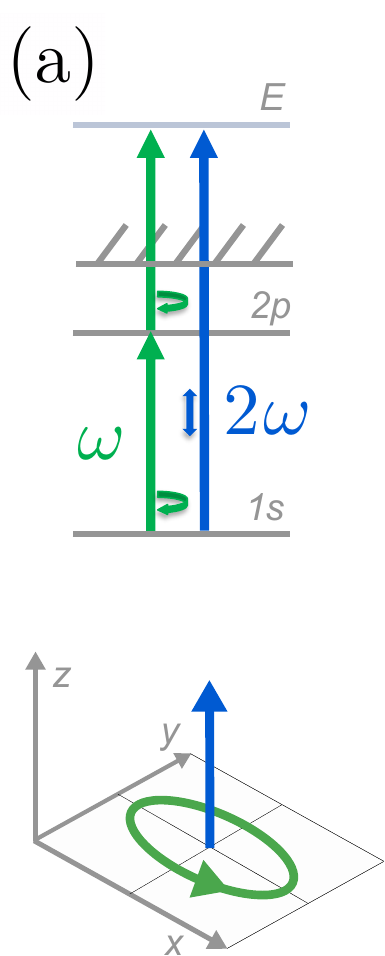}%
        }%
    &
    \begin{tabular}{c}
        \includegraphics[scale=1]{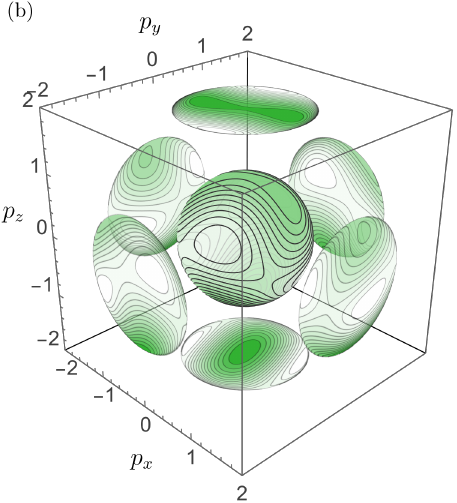}%
        \includegraphics[scale=1]{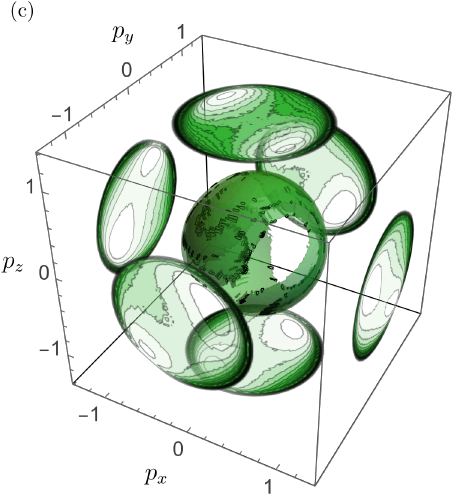}%
        \\[3mm]
        \includegraphics[scale=1]{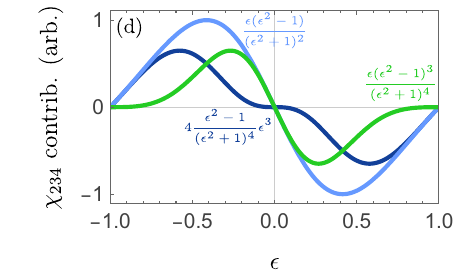}
        \includegraphics[scale=1]{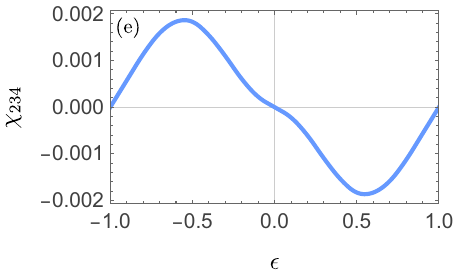}
    \end{tabular}
    \end{tabular}
    \caption{
    Photoionization of hydrogen by bichromatic chiral light.
    (a) Level scheme and field configuration.
    (b) Photoelectron momentum distribution $|\psi(\vbp)|^2$ obtained via perturbation theory, as per~\eqref{pt-final-wavefunction-main-text}, shown as a contour plot restricted to the sphere with radius $p_0$ (in atomic units), together with two-dimensional contour maps of the restriction of the distribution to each hemisphere.
    We show $|\psi(\vbp)|^2$ for the case when $c_\mathrm{s}(p)=0$, $c_\mathrm{p}(p)=c_\mathrm{d}(p)=1$, $\epsilon = 1/2$, $\varphi = \SI{45}{\degree}$, $E_{1,0}=E_{2,0}=1$.
    (c) Photoelectron momentum distribution obtained by numerical simulation of the TDSE, as described in Appendix~\ref{app-tdse-details}, for the case of $\epsilon = 0.6$.
    (d) Ellipticity dependence of the different possible contributions to $\chi_{234}$ for the perturbation-theory distribution, as derived in~\eqref{PT-chi234-final-result}
    (e) Ellipticity dependence of $\chi_{234}$ for the numerical TDSE simulation.
    }
    \label{fig:pt-figures}
\end{figure*}

In this section we illustrate how our chiral moments can be applied for this purpose.
To do this, we use the simplest example of photoionization driven by a chiral field: resonantly-enhanced two-photon ionization of atomic hydrogen driven by an elliptically-polarized fundamental, combined with a second harmonic with linear polarization orthogonal to the fundamental's plane of ellipticity,
a configuration which has been proposed in the infrared range for giant enantiosensitive optical responses through high-order harmonic generation~\cite{Ayuso2019};
we show a schematic in Fig.~\ref{fig:pt-figures}.
The resulting photoelectron momentum distribution is chiral due to the interference between the two ionization channels.

For this process, the final momentum-representation photoelectron wavefunction can be derived using perturbation theory, yielding
\begin{align}
\psi(\vbp)
& =
c_\mathrm{p}(p)
\:
\vbE_2
{\cdot}
\vbp
+
c_\mathrm{s}(p)
\:
\vbE_1
{\cdot}
\vbE_1 
+
c_\mathrm{d}(p)
\:
\proj{2}\vbE_1^{\otimes 2}
\tensordot
\vbp^{\otimes 2}
,
\label{pt-final-wavefunction-main-text}
\end{align}
for which we defer the details to Appendix~\ref{sec-appendix-perturbation-theory},
where $\vbE_1$ and $\vbE_2$ are the complex amplitudes of the fundamental and second-harmonic fields, 
and $c_\mathrm{s}(p)$, $c_\mathrm{p}(p)$ and $c_\mathrm{d}(p)$ are complex amplitudes describing the s-, p- and d-wave ionization channels.
The photoelectron distribution is then obtained as $\rho(\vbp) = |\psi(\vbp)|^2$ and its tensorial multipolar moments, which are easy to combine into a chiral moment, for which the natural choice is $\chi_{234}$.

For arbitrary polarizations of the fields, $\vbE_1$ and $\vbE_2$, it is possible (as we argue in more detail in Appendix~\ref{sec-appendix-perturbation-theory}) to obtain $\chi_{234}$ as a scalar polynomial in $\vbE_1$ and $\vbE_2$, and their conjugates, of mixed degree, involving terms of ninth and eleventh order.
These polynomials are nonlinear chiral correlation functions of the field, as introduced for synthetic chiral light in \cite{Ayuso2019}; in that notation, they would read as $h^{(9)}$ and $h^{(11)}$.
These relatively high orders are in contrast with the lower-order correlation function ($h^{(5)}$) which appears for this field configuration for perturbative nonlinear optical processes~\cite{Ayuso2019}, with the difference driven by the fact that $\rho(\vbp)$, as the square of the wavefunction, is a higher-order object than the wavefunction as used in the case of~$h^{(5)}$.

For the specific case of an elliptically-polarized fundamental and an out-of-plane linearly-polarized second harmonic, the chiral moment simplifies to an explicit analytical dependence on the ellipticity $\epsilon$ of the fundamental and the relative phase $\varphi$ between the two fields,
which we provide as Eq.~\eqref{PT-chi234-final-result} in the Appendices.
The ellipticity dependence is proportional to three terms given by
$
\frac{
  \epsilon^2 - 1
  }{
  (\epsilon^2 + 1)^4
  }
\epsilon^3
$,
$
\frac{
  \epsilon
  (\epsilon^2 - 1)
  }{
  (\epsilon^2 + 1)^2
  }
$,
and
$
\frac{
  \epsilon
  (\epsilon^2 - 1)^3
  }{
  (\epsilon^2 + 1)^4
  }
$,
with relative contributions determined by the relative amplitudes of the two fields as well as their spectra and the details of the atomic structure.
We show in Figure~\ref{fig:pt-figures} the shape of these ellipticity dependences.
The angular dependence of the photoelectron momentum distribution, $\rho(\vbp)$, is also clearly and graphically chiral;
we show an example in Figure~\ref{fig:pt-figures}.

In addition to the perturbation-theory calculation, we simulate this process through the numerical solution of the time-dependent Schr\"odinger equation (TDSE),
which provides a valuable test case of the application of our open-source software tools~\cite{pisanty-Chimera-2026} for extracting the chiral moments from raw data (whether from simulations or experiment).
We detail our TDSE simulations in Appendix~\ref{app-tdse-details}.

\section{Discussion}
\label{sec-discussion}
As we have seen, for both 
concrete distributions with direct links to experiment 
as well as for
a variety of model distributions,
our formalism provides intuitive, flexible and robust ways of tackling one of the thornier questions in chirality~-- assigning a sign to the handedness of a particular distribution.
This comes as a generalization and extension of existing multipole-moment based frameworks~\cite{Harris1999, Osipov1995, Neal2003, Hattne2011}, 
with a much clearer link to the relevant shapes that embody the chirality,
as well as 
an understanding of the role of the radial structure of the distribution~--
a possibility which has been recognized~\cite{Harris1999, Osipov1995, Hattne2011}, but not followed up on.
The ability to incorporate the radial features of the distribution allows
on the one hand,
more robust features, including with respect to noise as well as e.g.\ the precise location of the spatial origin,
and, more generally,
for a family of chiral measures with a wider scope, and thus less susceptible to the `blind spots' that result from the rubber-glove theorem~\cite{Walba1991, Weinberg2011, Fowler2005}.

\begin{figure}[t!]
    \centering
    \includegraphics[width=0.8\linewidth]{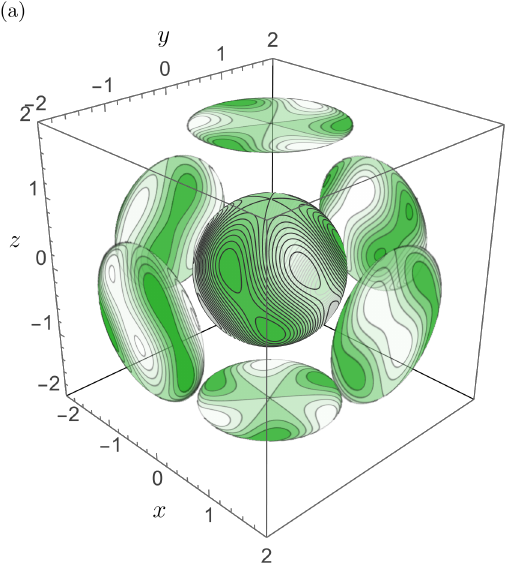}
    \includegraphics[width=0.8\linewidth]{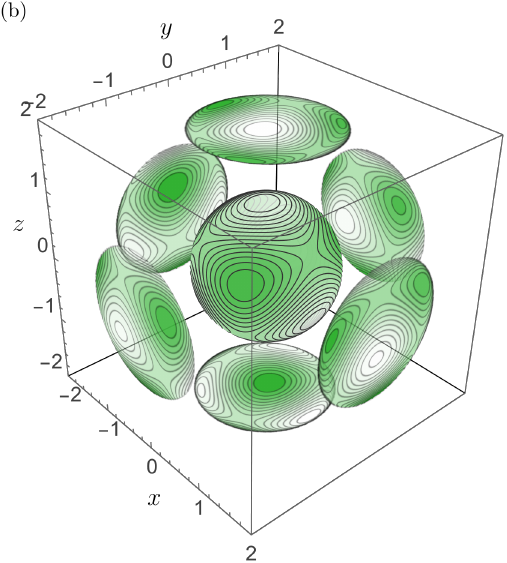}
    \caption{
    (a) Apparent `blind spot' distribution, $\rho_\mathrm{ABS}(\vbr)$, as described in Appendix~\ref{app-blind-spot-Y33-Y43}, plotted as in Fig.~\ref{fig:pt-figures}.
    (b) Purely octupolar `blind spot' distribution, $\rho_\mathrm{POBS}(\vbr)$, as described in Appendix~\ref{app-blind-spot-pure-octupole}.
    }
    \label{fig:blind-spots}
\end{figure}

That said, it is important to acknowledge that, even for the full generality of our chiral measures $\chi_{n_1 n_2 n_3}^{2q_1, 2q_2, 2q_3}$, some `blind spots' still remain~-- some of which can be folded into the tensor-cross-product formalism, and some of which cannot.

The simplest example of the former is a distribution of the form 
$\rho(\vbr) = \Re[Y_{33}(\vbr)+ e^{i\varphi} Y_{43}(\vbr)]$ 
constrained to a sphere, as shown in Figure~\ref{fig:blind-spots}(a), which we briefly explore in Appendix~\ref{app-blind-spot-Y33-Y43}.
For structures such as these, 
where only two different multipolar moments ($\boldsymbol{\mu}^{(3)}$ and $\boldsymbol{\mu}^{(4)}$) are nonzero,
a chiral measure based directly on triple tensor products will always vanish.%
\footnote{%
The general tensorial moments can be nonzero, such as e.g.\ $\tens{M}^{(5)} \propto \sym( \boldsymbol{\mu}^{(3)} \otimes \id)$, but they do not provide new information in this case.
}
Nevertheless, it is still possible to combine multiple cross products of $\boldsymbol{\mu}^{(3)}$ and $\boldsymbol{\mu}^{(4)}$, of different ranks, into a chirally-sensitive pseudoscalar,
which takes the form of a \textit{six}-point correlation function of $\rho(\vbr)$,
and which we provide in Appendix~\ref{app-blind-spot-Y33-Y43}.

An example of the latter~-- a distribution whose chirality cannot (currently) be quantified by our tensor-cross-product formalism~-- can be built as a chiral superposition of purely octupolar components ($Y_{3m}(\vbr)$) with different amplitudes for the different $m$ values, 
as shown in Figure~\ref{fig:blind-spots}(b),
and which we explore briefly in Appendix~\ref{app-blind-spot-pure-octupole}.
For such a distribution, only one tensorial multipolar moment, $\boldsymbol{\mu}^{(3)}$, is nonzero,%
\footnote{%
As above, other general tensorial moments can be nonzero, such as $\tens{M}^{(5)} \propto \sym( \boldsymbol{\mu}^{(3)} \otimes \id)$, but they do not provide new information.
}
and thus no tensor triple product is directly applicable.
This example thus presents a `blind spot' for our framework (or, at least, an open question), as well as for the existing chirality measures from the literature,
and, as such, appears as a distributional example of what is known in chiral measurements as `cryptochirality'~\cite{Mislow1976}: chiral structures whose chirality is difficult to quantify or to relate to experimental observables.

It is also useful to consider expanding the range of our framework to include spatial variation of chirality, as well as its dependence on the relevant length scales.
For those purposes, one readily-applicable approach is to introduce spatial filters (e.g. gaussians) to take local values of the tensor moments, which can then produce a relevant local gaussian-filtered chiral moment.
Even more locally, the behaviour of the distribution $\rho(\vbr)$ at a chosen point $\vbr_0$ can be understood purely in terms of its (partial) derivatives, which can then also be combined, as tensors, in the same way as the distribution's moments.
In component notation, that would read, for example, in the form
\begin{align}
\chi_{234}^\mathrm{(local)} (\vbr)
& =
\epsilon_{ijk}
\left(\partial_i\partial_l\rho \right)
\left(\partial_j\partial_m\partial_n\rho \right)
\left(\partial_k\partial_l\partial_m\partial_n\rho \right)
,
\end{align}
using the shorthand $\partial_i$ for $\frac{\partial}{\partial x_i}$.%
\footnote{
Alternatively, in invariant notation this takes a form mirroring the chiral moments:
$
\chi_{234}^\mathrm{(local)} (\vbr)
=
\left(
  \nabla^{\otimes 2} \rho
  \times
  \nabla^{\otimes 3} \rho
  \right)^{(4)}
\tensordot
\nabla^{\otimes 4} \rho
$.
}

On a more general perspective, not all relevant distributions need to be of a chiral shape in their own right, and often the chirality of an experimental result comes from the relationship between the observed distribution and the surrounding experiment (for example, forwards-backwards asymmetry with respect to a preferred axis).
This is particularly important for the chiral study of light-matter interactions (such as photoelectron circular dichroism~\cite{Powis2008, Sparling2022, Sparling2024, Sparling2025, Nahon2015, Lux2012, Lux2015, Powis1992}), 
where the driving light acts as a chiral reagent, providing an important spatial reference (e.g.\ a preferred axis) for the final observable.
For these sorts of cases, our formalism provides a natural framework, independent of the frame of reference, for incorporating these spatial features.
Thus, for example, the usual chiroptical observables will often have the form of tensor triple products, where one or more of the tensor factors is a light- or experiment-derived quantity,
such as 
the propagation direction light, $\vb k$, 
a polarization tensor of either quadratic ($\tens{P} = \left\langle \vb E(t)^{\otimes 2}\right\rangle$) or higher order (e.g.\ $\tens{T}^{(3)} = \left\langle \vb E(t)^{\otimes 3}\right\rangle$~\cite{Pisanty2019}),
or the spatial distribution of other chemical fragments obtained in a dissociative process~\cite{Pier2020, Berakdar1993, Viefhaus1996}.

Our formalism also extends naturally 
from structural chirality, which lives in space only, 
to dynamical chirality, which also includes the temporal dependence of the distribution.
For these cases, one can simply replace one or more of the tensorial-moment factors in~\eqref{chi-initial-definition} with a temporal derivative of such moments;
such a development would mirror the recent introduction of dynamical chirality in light~\cite{Ayuso2019, Ordonez2018, Ordonez2023}.
A similar development might be to extend our formalism to deal with vector-valued~\cite{Dryzun2011, Ordonez2023, Smirnova2022, Roos2026} (or even tensor-valued~\cite{Andrews2023, Hayami2018}) chiral distributions, in contrast to the scalar-valued ones we have considered so far.

A desirable broader application area for our formalism is to describe chiral optical responses in their full generality, 
from circular dichroism and optical rotation~\cite{Condon1937} to chiral Raman~\cite{Parchansky2014} and Rayleigh~\cite{Collins2019, McArthur2025} scattering and all the way up to extreme nonlinear optical processes~\cite{Ayuso2019, Khokhlova2022, Mayer2022, Mayer2024, Ayuso2022road, Ayuso2021, Ordonez2020, Ordonez2019i, Ordonez2019ii, Ordonez2018}.
Since chiral optical processes often involve tensor quantities of mixed symmetry~\cite{McArthur2025, Collins2019}, this would require generalizing our tensor triple product~\eqref{tensor-triple-product-definition} 
(as well as the tensor cross product~\eqref{tensor-cross-product-definition})
to allow for mixed-symmetry tensor factors,
using the connection to the Wigner $3j$ symbols (as explored in Appendix~\ref{app-multipolar-tensor-technical}) as a clear guide.

Already in its current generality, however, our work provides, as we have seen, a robust, flexible, intuitive, highly geometrical, physically-driven framework for understanding and quantifying the chirality of a wide variety of distributions.
The analytical toolset is matched by a corresponding open-source software package~\cite{pisanty-Chimera-2026}, which can deal with both symbolic computation as well as numerical simulation and experimental data.

\section*{Acknowledgements}
We are deeply grateful to the organizers and participants of the CUPUSL23 meeting in MPI-PKS, Dresden, where this work had its genesis, and particularly to Misha Ivanov and Olga Smirnova for inspiration and encouragement and to David Ayuso for stimulating discussions.

E.P.\ acknowledges Royal Society funding under URF\textbackslash{}\allowbreak{}R1\textbackslash{}\allowbreak{}211390.
N.M.\ acknowledges funding by the UK Research and Innovation (UKRI) under the UK government’s Horizon Europe funding guarantee [Grant No. EP/Z001390/1].
A.O.\ acknowledges funding from the Deutsche Forschungsgemeinschaft (DFG, German Research Foundation) 543760364.
A.L.\ acknowledges funding from the European Union under ERC, ULISSES, 101054696.
M.K.\ acknowledges Royal Society funding under URF\textbackslash{}\allowbreak{}R1\textbackslash{}\allowbreak{}231460.

E.P.\ dedicates this work to the memory of Prof.\ Eugenio Ley Koo.

\appendix

\renewcommand{\theequation}{\Alph{section}.\arabic{equation}}
\numberwithin{equation}{section}


\section{Definitions and properties}
\label{app-definitions-and-properties}

The central problem in constructing our chiral measures can be expressed as follows:
given three symmetric tensors, 
$\mathbf{A}^{(n_1)}$, $\mathbf{B}^{(n_2)}$ and $\mathbf{C}^{(n_3)}$,
with components
$A_{i_1\cdots i_{n_1}}$, $B_{i_1\cdots i_{n_2}}$ and $C_{i_1\cdots i_{n_3}}$,
we would like to form a single pseudoscalar~---
that is, a single scalar number, independent of the frame of reference,
and which changes sign under spatial inversion and reflections.

There is one clear way to do this, achievable via tensor contractions between the three tensor factors together with one added Levi-Civita tensor factor.
This is the tensor triple product, defined in Eq.~\eqref{tensor-triple-product-definition}, which we recall as
\begin{align}
\left( 
  \mathbf{A}^{(n_1)}
  \tensorcross
  \mathbf{B}^{(n_2)}
  \right)^{(n_3)}
\tensordot
\mathbf{C}^{(n_3)}
& =
\epsilon_{i_1 i_2 i_3}
A_{i_1 \mathbf{j} \mathbf{k}}
B_{i_2 \mathbf{j} \mathbf{l}}
C_{i_3 \mathbf{k} \mathbf{l}}
.
\nonumber
%
\end{align}
The structure is simple: the Levi-Civita factor is contracted, via each of its three indices, with one index each from 
$\mathbf{A}^{(n_1)}$, $\mathbf{B}^{(n_2)}$ and $\mathbf{C}^{(n_3)}$,
and these are then contracted with each other.
These contractions are taken through the repeated multi-indices
$\mathbf j = (j_1\cdots j_{m})$,
$\mathbf k = (k_1\cdots k_{n})$ and
$\mathbf l = (l_1\cdots l_{p})$
that appear shared between the three pairs of factors.
Finally, the lengths $m$, $n$ and $p$ of the shared multi-indices can be found by requiring that no indices remain free, which a simple calculation shows to require
$m=\frac12(n_1 + n_2 - n_3 -1)$,
$n=\frac12(n_1 + n_3 - n_2 -1)$ and 
$p=\frac12(n_2 + n_3 - n_1 -1)$.

As a quick note, the form \eqref{tensor-triple-product-definition} arises from the theory of isotropic tensors~\cite{Andrews1977, Jeffreys1973} as the single clear candidate for our pseudoscalar.
While other combinations are possible, they would involve internal contractions (i.e., tensor traces) between indices belonging to the same tensor factor;
as such, they would involve factors of, e.g., $\Tr(\tens{A}^{(n_1)})$, 
and can therefore be taken separately.
More concretely, the form \eqref{tensor-triple-product-definition} is the unique pseudoscalar with maximal connectivity between the indices of the three tensor factors.

As a quick note, our definition \eqref{tensor-triple-product-definition} for the tensor triple product relies on the fact that the three tensor factors are fully symmetric, 
which removes any concern about which indices are contracted together.
For our purposes, this is justified, as the moment tensors $\tens{M}^{(n)}$ and their multipolar counterparts $\boldsymbol{\mu}^{(\ell)}$ are all symmetric.
Nevertheless, the extension of this definition to tensors of mixed symmetry is an interesting (and, thus far, open) question, with clear applications to the pseudoscalars that appear in general chiroptical experiments.

Once the pseudoscalar~\eqref{tensor-triple-product-definition} has been defined, it provides us with a clear and unique definition of the tensor cross product.
In particular, we can see that Eq.~\eqref{tensor-triple-product-definition} defines a linear mapping
$
\tens{C}^{(n_3)} 
\mapsto 
\epsilon_{i_1 i_2 i_3}
A_{i_1 \mathbf{j} \mathbf{k}}
B_{i_2 \mathbf{j} \mathbf{l}}
C_{i_3 \mathbf{k} \mathbf{l}}
$,
which takes a symmetric complex-valued tensor and returns a scalar.
This mapping can always be interpreted as the full contraction with a separate fully-symmetric tensor, and this is what we define to be the tensor cross product:
\begin{align}
\left( 
  \mathbf{A}^{(n_1)}
  \tensorcross
  \mathbf{B}^{(n_2)}
  \right)^{(n_3)}_{i_3 \mathbf{k} \mathbf{l}}
& 
=
\sym_{i_3 \mathbf{k} \mathbf{l}}
\bigg(
  \epsilon_{i_1 i_2 i_3}
  A_{i_1 \mathbf{j} \mathbf{k}}
  B_{i_2 \mathbf{j} \mathbf{l}}
  \bigg)
\label{tensor-cross-product-definition}
\end{align}
%
where $\sym_{i_3 \mathbf{k} \mathbf{l}}$ denotes full symmetrisation over the remaining free indices, $i_3$, $\mathbf{k}$ and $\mathbf{l}$.

One clear route for getting more solid intuition into the meaning, behaviour, and handling of the tensor triple product, and the tensor cross product,
is to consider the scenario when each of the factors is a simple tensor power of a chosen vector, 
i.e.\ when
$\mathbf{A}^{(n_1)} = \vb a^{\otimes n_1}$, 
$\mathbf{B}^{(n_2)} = \vb b^{\otimes n_2}$ and 
$\mathbf{C}^{(n_3)} = \vb c^{\otimes n_3}$,
(in component notation $A^{(n_1)}_{i_1\cdots i_{n_1}} = a_{i_1} \cdots a_{i_{n_1}}$, and similarly for $\mathbf{B}^{(n_2)}$ and $\mathbf{C}^{(n_3)}$).
For this case, the Levi-Civita contraction turns into a simple vector triple product, $\left(\left( \vb a \times \vb b \right) \cdot \vb c \right)$,
while the pairwise tensor contractions become simple vector dot products,
resulting in a clean expression:
\begin{align}
\Big( 
  \vb a^{\otimes n_1} 
  \tensorcross
  &
  \vb b^{\otimes n_2}
  \Big)^{(n_3)}
\tensordot
\vb c^{\otimes n_3}
= 
\\ & \nonumber
\left(\left( \vb a \times \vb b \right) \cdot \vb c \right)
\left(\vb a \cdot \vb b \right)^{m}
\left(\vb b \cdot \vb c \right)^{n}
\left(\vb c \cdot \vb a \right)^{p}
.
\end{align}
Similarly, the tensor cross product for this case is 
\begin{align}
\left( 
  \vb a^{\otimes n_1}
  \tensorcross
  \vb b^{\otimes n_2}
  \right)^{(n_3)}
& =
\sym(
\left( \vb a \times \vb b \right) 
\left(\vb a \cdot \vb b \right)^{m}
\otimes 
\vb b^{\otimes n}
\otimes
\vb a^{\otimes p}
)
,
\end{align}
i.e., a symmetrized tensor product between the cross product $\vb a \times \vb b $ and the tensor powers $\vb b^{\otimes n}$ and $\vb a^{\otimes p}$, all multiplied by a power of their dot product, $\left(\vb a \cdot \vb b \right)^{m}$.

For simple cases, it is often fairly feasible to calculate the tensor cross product directly, at least when the tensor ranks involved are low.
For higher ranks, and for more complex cases as well as for numerical data, however, this is impractical. 
For those settings, we have implemented the tensor cross product in the Wolfram Language as the open-source software package \texttt{Chimera} available as Ref.~\cite{pisanty-Chimera-2026}.

To make this section complete, we should point out that this formalism also admits a clean extension for the case of even parity (i.e.\ when the tensor factors need to be combined into a single scalar, rather than a pseudoscalar), in which case we write
\begin{align}
\left( 
  \mathbf{A}^{(n_1)}
  \tensorcross
  \mathbf{B}^{(n_2)}
  \right)^{(n_3)}
\tensordot
\mathbf{C}^{(n_3)}
& =
A_{\mathbf{j} \mathbf{k}}
B_{\mathbf{j} \mathbf{l}}
C_{\mathbf{k} \mathbf{l}}
,
\label{tensor-triple-product-even-parity-def}
\end{align}
omitting the Levi-Civita factor. 
For this case, the multi-indices are still denoted
$\mathbf j = (j_1\cdots j_{m})$,
$\mathbf k = (k_1\cdots k_{n})$ and
$\mathbf l = (l_1\cdots l_{p})$,
though their lengths now take the values
$m=\frac12(n_1 + n_2 - n_3)$,
$n=\frac12(n_1 + n_3 - n_2)$ and 
$p=\frac12(n_2 + n_3 - n_1)$.
We will use this even-parity product to combine three pseudotensors in Appendix~\ref{app-blind-spot-Y33-Y43}, as a route to dealing with the apparent `blind spot' mentioned in the Discussion.

\section{Explicit forms for the traceless kernels}
\label{app-explicit-forms}

As mentioned in the main text, our chiral moments 
$
\chi_{n_1 n_2 n_3}
=
\left( 
  \mathbf{M}^{(n_1)}
  \tensorcross
  \mathbf{M}^{(n_2)}
  \right)^{(n_3)}
\tensordot
\mathbf{M}^{(n_3)}
$
can be re-ex\-pres\-sed as a triple integral, given by equation~\eqref{chiral-moment-as-triple-integral}, which we recall as
\begin{align*}
\chi_{n_1 n_2 n_3}
& =
\iiint
\d\vbr_1
\d\vbr_2
\d\vbr_3
\rho(\vbr_1)
\rho(\vbr_2)
\rho(\vbr_3)
\\ & \qquad \qquad \quad \nonumber
\left( 
  \vbr_1^{\otimes n_1}
  \tensorcross
  \vbr_2^{\otimes n_2}
  \right)^{(n_3)}
\tensordot
\vbr_3^{\otimes n_3}
\\
& =
\iiint
\d\vbr_1
\d\vbr_2
\d\vbr_3
\rho(\vbr_1)
\rho(\vbr_2)
\rho(\vbr_3)
%
%
\\ & \qquad \quad \nonumber
\left[
  (\vbr_1 \times \vbr_2) \cdot \vbr_3 
  \right]
(\vbr_1 \cdot \vbr_2)^{m}
(\vbr_1 \cdot \vbr_3)^{n}
(\vbr_2 \cdot \vbr_3)^{p}
.
\end{align*}
Similarly, we defined the traceless chiral moments as
$
h_{\ell_1 \ell_2 \ell_3}
=
\left( 
  \boldsymbol{\mu}^{(\ell_1)}
  \tensorcross
  \boldsymbol{\mu}^{(\ell_2)}
  \right)^{(\ell_3)}
\tensordot
\boldsymbol{\mu}^{(\ell_3)}
$,
in terms of the (traceless) tensorial multipolar moments
$
\boldsymbol{\mu}^{(\ell)} = \hat{\Pi}_\ell \mathbf{M}^{(\ell)}
$,
and, for these traceless chiral moments, a similar argument applies:
the three integrals in the $\boldsymbol{\mu}^{(\ell)}$ can be expanded out and brought together, giving an expression for the traceless chiral moment of the form
\begin{align}
h_{\ell_1 \ell_2 \ell_3}
& =
\iiint
\d\vbr_1
\d\vbr_2
\d\vbr_3
\:
\rho(\vbr_1)
\rho(\vbr_2)
\rho(\vbr_3)
\\ & \qquad \qquad \quad \nonumber
\left( 
  \hat{\Pi}_{\ell_1}
  \vbr_1^{\otimes \ell_1}
  \tensorcross
  \hat{\Pi}_{\ell_2}
  \vbr_2^{\otimes \ell_2}
  \right)^{(\ell_3)}
\tensordot
\hat{\Pi}_{\ell_3}
\vbr_3^{\otimes \ell_3}
\\ &
\!\!\!\!\!\!\!\!
=
\nonumber 
\iiint
\d\vbr_1
\d\vbr_2
\d\vbr_3
\:
\rho(\vbr_1)
\rho(\vbr_2)
\rho(\vbr_3)
H_{\ell_1 \ell_2 \ell_3}(\vbr_1, \vbr_2, \vbr_3)
.
\end{align}
As for the `unabridged' chiral moments, this takes the form of a three-point correlation function, with three copies of the density $\rho(\vbr)$ multiplied against a correlated kernel,
\begin{align}
H_{\ell_1 \ell_2 \ell_3}(\vbr_1, \vbr_2, \vbr_3)
& =
\left( 
  \hat{\Pi}_{\ell_1}
  \vbr_1^{\otimes \ell_1}
  \tensorcross
  \hat{\Pi}_{\ell_2}
  \vbr_2^{\otimes \ell_2}
  \right)^{(\ell_3)}
\tensordot
\hat{\Pi}_{\ell_3}
\vbr_3^{\otimes \ell_3}
,
\end{align}
which is again a (pseudo)scalar homogeneous polynomial in the Cartesian components of $\vbr_1$, $\vbr_2$ and $\vbr_3$, of respective orders $\ell_1$, $\ell_2$ and $\ell_3$, 
though now it is fully traceless, in the sense that
it is a harmonic function of each of its variables, i.e.,
\begin{align}
\nabla^2_{\vbr_i}
H_{\ell_1 \ell_2 \ell_3}(\vbr_1, \vbr_2, \vbr_3)
=
0,
\end{align}
for $i=1,2,3$.%
\footnote{
This tracelessness also manifests as the fact that
a spherical integral of $H_{\ell_1 \ell_2 \ell_3}(\vbr_1, \vbr_2, \vbr_3)$ against any spherical or solid harmonic $S_{\ell,m}(\vbr_1)$ of degree $\ell<\ell_1$ will vanish, i.e.\ 
%
$
\int_{r_1=1} 
H_{\ell_1 \ell_2 \ell_3}(\vbr_1, \allowbreak{} \vbr_2, \vbr_3)
S_{\ell,m}(\vbr_1)
\d\Omega_1
=
0
$%
,
and similarly for integrals over $\vbr_2$ and~$\vbr_3$.
}

These polynomials must be constructed out of dot products and norms of the $\vbr_i$, and, as such, they are relatively simple to construct~-- and, once constructed, their explicit forms are simple to verify via computer algebra.
These explicit forms read, for the first few cases, as follows:
\begin{widetext}
\begin{subequations}%
\allowdisplaybreaks[1]%
\begin{align}%
H_{111}(\vbr_1, \vbr_2, \vbr_3)
& =
(\vbr_1 \times \vbr_2)\cdot \vbr_3
%
%
,
\\
H_{122}(\vbr_1, \vbr_2, \vbr_3)
& =
\big((\vbr_1 \times \vbr_2)\cdot \vbr_3\big)
(\vbr_2\cdot\vbr_3)
\vphantom{\bigg]}
,
\\
H_{223}(\vbr_1, \vbr_2, \vbr_3)
& =
\big((\vbr_1 \times \vbr_2)\cdot \vbr_3\big)
\left[
  (\vbr_1\cdot\vbr_3) (\vbr_2\cdot\vbr_3)
  -\tfrac15 (\vbr_1\cdot\vbr_2) r_3^2
  \right]
,
\\
H_{133}(\vbr_1, \vbr_2, \vbr_3)
& =
\big((\vbr_1 \times \vbr_2)\cdot \vbr_3\big)
\left[
  (\vbr_2\cdot\vbr_3)^2
  -\tfrac15 r_2^2 r_3^2
  \right]
,
\\
H_{144}(\vbr_1, \vbr_2, \vbr_3)
& =
\big((\vbr_1 \times \vbr_2)\cdot \vbr_3\big)
(\vbr_2\cdot\vbr_3)
\left[
  (\vbr_2\cdot\vbr_3)^2
  -\tfrac37 r_2^2 r_3^2
  \right]
,
\\
H_{234}(\vbr_1, \vbr_2, \vbr_3)
& =
\big((\vbr_1 \times \vbr_2)\cdot \vbr_3\big)
\left[
  (\vbr_1\cdot\vbr_3)(\vbr_2\cdot\vbr_3)^2
  -\tfrac27 (\vbr_1\cdot\vbr_2)(\vbr_2\cdot\vbr_3) r_3^2
  -\tfrac17 (\vbr_1\cdot\vbr_3) r_2^2 r_3^2
  \right]
,
\\
H_{333}(\vbr_1, \vbr_2, \vbr_3)
& =
\big((\vbr_1 \times \vbr_2)\cdot \vbr_3\big)
\left[
  (\vbr_1\cdot\vbr_2)(\vbr_2\cdot\vbr_3)(\vbr_3\cdot\vbr_1)
  - \tfrac15 
      \left(
          r_1^2 (\vbr_2{\cdot}\vbr_3)^2
        {+} r_2^2 (\vbr_3{\cdot}\vbr_1)^2
        {+} r_3^2 (\vbr_1{\cdot}\vbr_2)^2
      \right)
  +\tfrac{2}{25} r_1^2 r_2^2 r_3^2
  \right]
.
\end{align}%
\end{subequations}%
\end{widetext}%
The specific coefficients involved (e.g. $\tfrac15$, $\tfrac37$, $\tfrac17$) are ultimately determined by the specific multipolar projectors~$\proj{\ell}$ involved 
(and, specifically, by the coefficients $c_{n,\ell}$ and $b_{\ell,m}$ we will develop in more detail in section~\ref{app-trace-removal-projector-properties} below).
The normalization is governed by requiring the leading term (i.e.\ the fully-connected contraction, leading to the combination that appears in \eqref{chiral-moment-as-triple-integral}) to have unit coefficient.

In more formal terms, the polynomials $H_{\ell_1 \ell_2 \ell_3}$ are known as the scalar `tripolar' spherical harmonics~\cite[\S5.16.2]{Varshalovich1988}, 
and they have seen some use in both atomic, molecular and optical physics~\cite{Manakov1996, Manakov1998, Manakov2002, Berakdar2004, DaPieve2007, Siminovitch2008, Ippolitov1985, McClelland1979, Barnwell1983, Malcherek1997}
(therein sometimes termed `triple product of degree zero'~\cite{Fano1959}),
as well as in nuclear spectroscopy~\cite{Biedenharn1960}
and cosmology~\cite{Szapudi2004, Borisennko2006, Joshi2010, Shiraishi2017, Shiraishi2021, Byun2023} (where they are sometimes abbreviated as `TripoSH'~\cite{Shiraishi2017, Shiraishi2021, Byun2023}),
though their detailed properties do not seem to have received much attention in the literature~\cite{Biedenharn1960, Szapudi2004}.
These polynomials also generalize to `polypolar'~\cite{Shiraishi2017} or `$n$-polar'~\cite{Malcherek1997} harmonics, which can accommodate arbitrary numbers of positions 
(and, thus, correspond to $n$-point correlation functions of $\rho(\vbr)$~\cite{Cahn2023test}),
thereby embodying the higher-order combination required to deal with the apparent `blind spot' as described in section~\ref{app-blind-spot-Y33-Y43} below.

Finally, for completeness, we provide here the corresponding even-parity polynomials.
\begin{widetext}
\begin{subequations}%
\allowdisplaybreaks[1]%
\begin{align}%
H_{112}(\vbr_1, \vbr_2, \vbr_3)
& =
(\vbr_1\cdot\vbr_3) (\vbr_2\cdot\vbr_3)
-\tfrac13 (\vbr_1\cdot\vbr_2) r_3^2
,
\\
H_{123}(\vbr_1, \vbr_2, \vbr_3)
& =
(\vbr_1\cdot\vbr_3) (\vbr_2\cdot\vbr_3)^2
-\tfrac25 (\vbr_1\cdot\vbr_2)(\vbr_2\cdot\vbr_3) r_3^2
-\tfrac15 (\vbr_1\cdot\vbr_3) r_2^2 r_3^2
,
\\
H_{222}(\vbr_1, \vbr_2, \vbr_3)
& =
\tfrac13 (\vbr_1\cdot\vbr_3) (\vbr_2\cdot\vbr_3) (\vbr_3\cdot\vbr_1)
-\tfrac19 r_1^2 r_2^2 r_3^2
+\tfrac13 \left((\vbr_1 \times \vbr_2)\cdot \vbr_3\right)^2
,
\\
H_{134}(\vbr_1, \vbr_2, \vbr_3)
& =
(\vbr_1\cdot\vbr_3) (\vbr_2\cdot\vbr_3)^3
-\tfrac37 (\vbr_1\cdot\vbr_3) (\vbr_2\cdot\vbr_3) r_2^2 r_3^2
-\tfrac37 (\vbr_1\cdot\vbr_2) (\vbr_2\cdot\vbr_3)^2 r_3^2
+\tfrac{3}{35} (\vbr_1\cdot\vbr_2) r_2^2 r_3^4
,
\\
H_{224}(\vbr_1, \vbr_2, \vbr_3)
& =
(\vbr_1\cdot\vbr_3)^2 (\vbr_2\cdot\vbr_3)^2
-\tfrac47 (\vbr_1\cdot\vbr_2) (\vbr_2\cdot\vbr_3)(\vbr_3\cdot\vbr_1) r_3^2
+\tfrac{2}{35} (\vbr_1\cdot\vbr_2)^2 r_3^4
\nonumber \\ & \qquad \qquad
-\tfrac17 (\vbr_1\cdot\vbr_3)^2 r_2^2 r_3^2
-\tfrac17 (\vbr_2\cdot\vbr_3)^2 r_1^2 r_3^2
+\tfrac{1}{35} r_1^2 r_2^2 r_3^4
,
\\
H_{233}(\vbr_1, \vbr_2, \vbr_3)
& =
(\vbr_1\cdot\vbr_2)(\vbr_1\cdot\vbr_3)(\vbr_2\cdot\vbr_3)^2
-\tfrac25 (\vbr_2\cdot\vbr_3) 
\left(
  (\vbr_1\cdot\vbr_3)^2 r_2^2
  + (\vbr_1\cdot\vbr_2)^2 r_3^2
  \right)
+\tfrac{1}{25} (\vbr_1\cdot\vbr_2) (\vbr_1\cdot\vbr_3) r_2^2 r_3^2
\nonumber \\ & \qquad \qquad
-\tfrac13 (\vbr_2\cdot\vbr_3)^3 r_1^2
+\tfrac{7}{25} (\vbr_2\cdot\vbr_3) r_1^2 r_2^2 r_3^2
.
\end{align}%
\end{subequations}%
\end{widetext}%

Here it is important to note that the scalar triple product, $(\vbr_1 \times \vbr_2)\cdot \vbr_3$, makes occasional appearances, such as in $H_{222}(\vbr_1, \vbr_2, \vbr_3)$.
In the achiral tripolar harmonics, the scalar triple product always appears squared, in which case it can always be written as a function of the pairwise dot products, $\vbr_i\cdot\vbr_j$, as the square is equal to the determinant of the Gram matrix:
\begin{equation}
\left((\vbr_1 \times \vbr_2)\cdot \vbr_3\right)^2
=
\det\mathopen{}\left[
  \left(\vbr_i\cdot\vbr_j\right)_{ij}
  \right]
.
\end{equation}

\section{Tensor theory}
\label{sec-appendix-tensor-technical}

Our definition of the tensorial multipolar moment 
$\boldsymbol{\mu}^{(\ell)} = \hat{\Pi}_\ell \mathbf{M}^{(\ell)}$
from the `unabridged' or `implicit' tensorial moment 
$\mathbf{M}^{(\ell)}$
hinges, crucially, on the trace-removal projector 
$\hat{\Pi}_\ell$
(also known as the `de-tracer' operator~\cite{Coles2020, Applequist1989, Applequist2002, Shanker2007}).
This Appendix summarizes its core properties, presents a general definition, and provides some useful properties.

\subsection{The trace-removal projector: intuition}
\label{app-trace-removal-projector-intuition}
The core distinction between the `unabridged' and the traceless moments is perhaps most familiar in the case of rank $n=2$, where the tensorial moment components read simply
\begin{align}
M^{(2)}_{ij}
=
\int
x_i x_j
\rho(\vbr)
\d\vbr
,
\qquad \qquad \quad 
\end{align}
whereas it is typically desirable to define the multipolar tensor moments having components
\begin{align}
\mu^{(2)}_{ij}
=
\int
\left(
  x_i x_j
  - \frac13 r^2 \delta_{ij}
  \right)
\rho(\vbr)
\d\vbr
,
\end{align}
as the component-notation reading of $\boldsymbol{\mu}^{(2)} = \proj{2} \tens{M}^{(2)}$.
This is done because it allows us to separate the `shape' information (contained in $\mu^{(2)}_{ij}$) from the rotationally-invariant trace $
\mu^{(0,2)} 
= \Tr\mathopen{}\left(\mathbf{M}^{(2)}\right) 
= M^{(2)}_{ii}
= \int r^2 \rho(\vbr)\d\vbr
$,
which describes the width of the distribution.
Thus, we decompose the tensorial moment as
\begin{align}
M^{(2)}_{ij}
=
\mu^{(2)}_{ij}
+
\tfrac13
\mu^{(0,2)} 
\delta_{ij}
,
\label{eq-app-M2-mu2}
\end{align}
and (since the Kronecker delta $\delta_{ij}$ is invariant under rotations) each of those two terms forms an isolated subspace under arbitrary rotations
(or, in the language of group representation theory~\cite{Hall2004}, an irreducible representation~\cite{Jerphagnon1978, Zou2001, Mane2016}).

For higher ranks, a similar separation holds.
Thus, for example, at rank $n=3$, we decompose the full tensor moment as
\begin{align}
M^{(3)}_{ijk}
=
\mu^{(3)}_{ijk}
+
\tfrac{3}{5}
\sym_{ijk}
\mu^{(1,2)}_i
\delta_{jk}
\end{align}
to separate the octupole tensor moment $\mu^{(3)}_{ijk}$ from 
$
\mu^{(1,2)}_i
=
\int x_i r^2 \rho(\vbr)\d\vbr
$,
the dipole moment of the distribution $r^2\rho(\vbr)$
(known in nuclear physics as the Schiff moment~\cite{Schiff1963,Flambaum2020}).
That said, from rank $n=4$ onwards, there are generally multiple terms that come into play, and the decomposition
\begin{align}
M^{(4)}_{ijkl}
=
\mu^{(4)}_{ijkl}
+
\tfrac{3}{5}
\sym_{ijkl}
\mu^{(2,2)}_{ij}
\delta_{kl}
+
\tfrac{6}{7}
\mu^{(0,4)} 
\sym_{ijkl}
\delta_{ij}
\delta_{kl}
,
\label{eq-app-M4-mus}
\end{align}
separates out
$\mu^{(2,2)}_{ij}$, the octupole moment tensor of the distribution $r^2\rho(\vbr)$,
as well as 
$\mu^{(0,4)}$, the rotationally-symmetric (monopole) average of the distribution $r^4\rho(\vbr)$,
since those two contributions form isolated subspaces that do not mix under arbitrary rotations.

The description of the general idea thus far, however, ignores the role of the fractional coefficients $\tfrac13$, $\tfrac35$ and $\tfrac67$ in Eqs.~\eqref{eq-app-M2-mu2} and~\eqref{eq-app-M4-mus}.
In essence, these are required to account for the various ways in which indices can be combined when e.g.\ 
recovering $\mu^{(2,2)}_{ij}$ from $M^{(4)}_{ijkl}$
by taking the trace of the latter, $M^{(4)}_{ijkk}$.
In the following section we will derive a general formula for these coefficients.

\subsection{The trace-removal projector: properties}
\label{app-trace-removal-projector-properties}
\begin{figure}[t]
    \centering
    \includegraphics[width=0.80\linewidth]{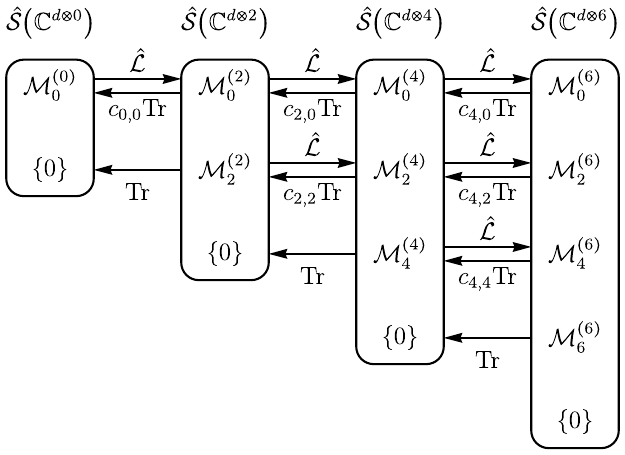}\\[2mm]
    \includegraphics[width=0.80\linewidth]{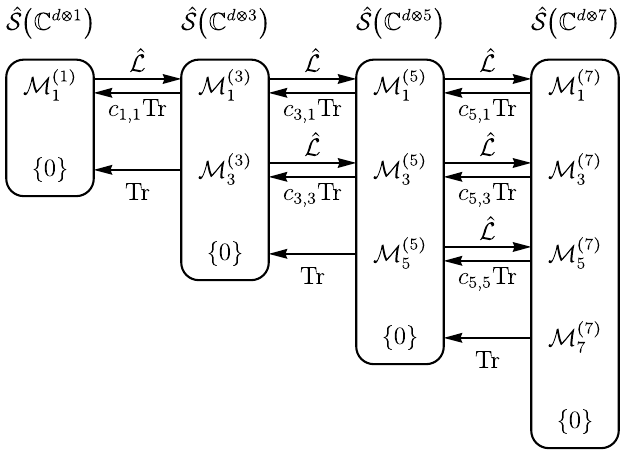}
    \caption{
    Schematic of the decomposition of $\symtens{n}$ as a direct sum of irreducible representations $\irrep{n}{\ell}$, and the links between the different subspaces via the trace and the tensor lift operator~$\lift$,
    for even parity (top) and odd parity (bottom).
    }
    \label{fig:lifts-and-traces}
\end{figure}

In more formal terms, we work in the fully-symmetric tensor space of rank $n$,
$\symtens{n}$, in dimension $d=3$.%
\footnote{Keeping $d$ symbolic will help keep the arithmetic clearer later on.}
We then decompose this tensor space as a direct sum of irreducible $\SO{3}$ representations with angular momentum number $\ell$, 
$
\symtens{n}
=
\bigoplus_{\ell=0}^n
\irrep{n}{\ell}
$,
with $\ell$ restricted to the same parity as $n$ in all summations in this section;
we illustrate this decomposition in Figure~\ref{fig:lifts-and-traces}.
We then define $\projj{n}{\ell}$ as the orthogonal projector onto $\irrep{n}{\ell}$, so that for any $\tens{A}^{(n)} \in \symtens{n}$ we will have
\begin{align}
\tens{A}^{(n)}
=
\sum_{\ell=0}^n
\projj{n}{\ell}
\tens{A}^{(n)}
.
\end{align}

We now need to establish ways to connect tensor spaces of different ranks.
In the `downward' direction, we already know the clear link:
it is simply the trace operator, $\Tr: \symtens{n} \to \symtens{(n-2)} $.
In the upward direction, the overall idea is as we used in Eq.~\eqref{eq-app-M4-mus}: 
multiply by the isotropic tensor $\mathbb{I}$ (whose components are the Kronecker delta, $\delta_{ij}$), 
in a symmetrized way;
in other words, we define $\lift: \symtens{n} \to \symtens{(n+2)}$ via
\begin{align}
\lift(\tens{A}^{(n)})
=
\sym(
  \tens{A}^{(n)}
  \otimes \mathbb{I}
  )
.
\end{align}

Both of these linking operations are invariant under $\SO{3}$ rotations, which implies that they will map the irreducible representation
$\irrep{n}{\ell}$
to its `neighbours', $\irrep{n\pm2}{\ell}$, 
as shown in Fig.~\ref{fig:lifts-and-traces},
crucially, preserving the angular momentum number $\ell$.
Moreover, since the $\irrep{n}{\ell}$ are irreducible representations,
Schur's lemma~\cite{Hall2004} implies that the composition $\Tr\circ\lift$ is a multiple of the identity,
or, in other words,
that the inverse of the tensor lift $\lift$ on $\irrep{n}{\ell}$ is $c_{n,\ell}\Tr$, 
as indicated in Fig.~\ref{fig:lifts-and-traces},
for some constant $c_{n,\ell}$.
These $c_{n,\ell}$ are the coefficients $\tfrac13$, $\tfrac35$ and $\tfrac67$ from Eqs.~\eqref{eq-app-M2-mu2} and~\eqref{eq-app-M4-mus}.

The values of the $c_{n,\ell}$ coefficients can be calculated via a recursive argument, built on the combined action of the trace and tensor lift operators.
This can be shown, for an arbitrary tensor $\tens{A} \in \symtens{n}$, to be given by
\begin{align}
\label{trace-of-sym-basic}
\Tr\mathopen{}\left[\sym(\tens{A}\otimes\id)\right]
& =
\frac{
  2(2n+d)
  }{
  (n+2)(n+1)
  }
\tens{A}
\\ \nonumber & \qquad 
+
\frac{
  n(n-1)
  }{
  (n+2)(n+1)
  }
\sym(\Tr(\tens{A})\otimes\id)
,
\end{align}
i.e., as a sum of 
our original tensor $\tens{A}$,
plus the `opposite' combination of the trace and lift operators, $\lift \circ \Tr$.
A recursive argument based on~\eqref{trace-of-sym-basic}, 
the proof of which we defer to section~\ref{app-trace-removal-projector-proofs},
then provides the desired coefficients.

The result of that calculation is the following pairs of operators and their inverses:
\begin{itemize}

\item
The inverse of $\lift$ on $\irrep{\ell+2m}{\ell}$ is $c_{\ell+2m,\ell}\Tr$, with coefficient
\begin{align}
c_{\ell+2m,\ell}
& =
\frac{
  (\ell+2m+2)(\ell+2m+1)
  }{
  2(m+1) (2\ell+2m+d)
  }
.
\end{align}

\item
The inverse of the $m$-fold lift $\lift^m$ on $\irrep{\ell}{\ell}$ is $b_{\ell,m}\Tr$, with coefficient
\begin{align}
b_{\ell,m}
& =
c_{\ell+2(m-1),\ell}
\cdots
c_{\ell,\ell}
\nonumber \\ & 
=
\frac{
  (\ell+2m)!
  }{
  2^m m!
  \ell!
  }
\frac{
  (2\ell-2+d)!!
  }{
  (2\ell+2(m-1)+d)!!
  }
,
\label{blm-result}
\end{align}
where $n!! = n\cdot (n-2)!!$ is the double factorial of $n$.

\end{itemize}
Alternatively, these two operator-inverse pairs can be rephrased as follows:
\begin{itemize}

\item
The inverse of $\lift$ on $\irrep{n}{\ell}$ is $c_{n,\ell}\Tr$, with coefficient
\begin{align}
c_{n,\ell}
& =
\frac{
  (n+2)(n+1)
  }{
  (n+2-\ell) (n+\ell+d)
  }
.
\end{align}

\item 
Starting on an arbitrary representation $\irrep{n}{\ell}$, 
the $m=\frac{n-\ell}{2}$-fold trace $b_{\ell,\frac{n-\ell}{2}} \Tr^{\frac{n-\ell}{2}}$ will map $\irrep{n}{\ell}$ onto $\irrep{\ell}{\ell}$, 
with inverse $\lift^{\frac{n-\ell}{2}}$, 
and with coefficient
\begin{align}
b_{\ell,\frac{n-\ell}{2}}
& =
\frac{
  n!
  }{
  2^{\frac{n-\ell}{2}} (\tfrac{n-\ell}{2})!
  \ell!
  }
\frac{
  (2\ell-2+d)!!
  }{
  (n+\ell-2+d)!!
  }
.
\label{blnl-result}
\end{align}

\end{itemize}

\subsection{The multipolar projectors \texorpdfstring{$\projj{n}{\ell}$}{Π(n,l)}}
\label{app-multipolar-projector-definition}

With this algebra in hand, it is now easy to provide an explicit definition for the trace-removal projectors $\projj{n}{\ell}$.
This is also done recursively, starting with the lowest angular momentum numbers $\ell=0$ and $\ell=1$, where we have
\begin{align}
\projj{n}{0} \tens{A} 
& = 
b_{0,\frac n2}\lift^{\frac n2}(\Tr^{\frac n2}(\tens{A}))
\nonumber \\ 
\projj{n}{1} \tens{A} 
& = 
b_{1,\frac{n-1}{2}}\lift^{\frac{n-1}{2}}(\Tr^{\frac{n-1}{2}}(\tens{A}))
,
\end{align}
since the trace operator, iterated as many times as will fit, will eliminate the contributions from $\ell>0$ (resp. $\ell>1$) representations.
For arbitrary $\ell$, the projector $\projj{n}{\ell}$ is defined as
\begin{align}
\projj{n}{\ell}\tens{A} 
& = 
b_{\ell,\frac{n-\ell}{2}}
\lift^{\frac{n-\ell}{2}}\mathopen{}\left(
  \Tr^{\frac {n-\ell}{2}}\mathopen{}\left(
    \tens{A}
    -
    \sum_{\ell'<\ell}
      \projj{n}{\ell'}
      \tens{A}
    \right)
  \right)
\label{pi-n-ell-definition}
\nonumber \\ & = 
b_{\ell,\frac{n-\ell}{2}}
\lift^{\frac{n-\ell}{2}}\mathopen{}\left(
  \Tr^{\frac {n-\ell}{2}}\mathopen{}\left(\tens{A}\right)
  -
  \sum_{\ell'<\ell}
    \projj{\ell}{\ell'}
    \Tr^{\frac{n-\ell}{2}}\mathopen{}\left(\tens{A}\right)
  \right)
.
\end{align}
i.e.,
first removing the representations with $\ell'<\ell$ using explicit projectors $\projj{n}{\ell'}$, 
and then removing the representations with $\ell'>\ell$ through a suitable combination of the trace and lift operators.
(This can then be simplified slightly by performing the $\ell'<\ell$ removal \textit{after} the $\frac{n-\ell}{2}$-fold trace, as in the second line of \eqref{pi-n-ell-definition}, i.e., while the tensor is momentarily living in $\symtens{\ell}$, which is smaller and thus more efficient.)

Finally, we turn briefly to some useful algebraic ways to frame these relationships. 
As mentioned earlier, the iterated lift
iterated lift 
$\lift^{\frac{n-\ell}{2}}$
is the inverse of the iterated trace 
$b_{\ell,\frac{n-\ell}{2}} \Tr^{\frac{n-\ell}{2}}$ 
acting on $\irrep{n}{\ell}$, so we can write
\begin{align}
\lift^{\frac{n-\ell}{2}}(
  b_{\ell,\frac{n-\ell}{2}} \Tr^{\frac{n-\ell}{2}}(
    \projj{n}{\ell}
    \tens{A}
    )
  )
& = 
\projj{n}{\ell}
\tens{A}
.
\end{align}
Moreover, it is also possible to shift where the multipolar projection happens, so that we get
\begin{align}
b_{\ell,\frac{n-\ell}{2}} 
\lift^{\frac{n-\ell}{2}}(
  \projj{\ell}{\ell}
  \Tr^{\frac{n-\ell}{2}}(
    \tens{A}
    )
  )
& = 
\projj{n}{\ell}
\tens{A}
,
\label{projj-as-trace-proj-lift}
\end{align}
and by applying the iterated trace once again, we can rephrase this as the operator commutation relation
\begin{align}
\projj{\ell}{\ell}
\Tr^{\frac{n-\ell}{2}}(
  \tens{A}
  )
& = 
\Tr^{\frac{n-\ell}{2}}(
  \projj{n}{\ell}
  \tens{A}
  )
.
\end{align}
Similar relationships, such as 
\begin{align}
\projj{n}{\ell}
\lift^{\frac{n-\ell}{2}}(\tens{A})
& =
\lift^{\frac{n-\ell}{2}}(
  \projj{\ell}{\ell}
  \tens{A}
  )
\end{align}
for $\tens{A}\in \symtens{\ell}$, can be derived through equivalent formulations.

\subsection{Extra tensor theory}
\label{app-multipolar-tensor-technical}
In this section we turn to further details of the tensorial multipolar moments, 
in order to provide a more fleshed-out theory,
but, most importantly, to show the key links between the tensorial version of the multipolar moments $\boldsymbol{\mu}^{(\ell)}$ and the more well-known spherical multipolar moments $M_{\ell m}$ which come from the standard integration of the form 
$
\int
r^\ell
Y_{\ell m}(\theta,\phi)
\rho(\vbr)
\d\vbr
$,
as well as to show a crucial link between our newly-defined triple tensor product and the standard Wigner $3j$ symbols.

We begin with the (traceless) tensorial multipolar moments, defined in Eq.~\eqref{mu-ell-definition}, which we recall as
\begin{align}
\boldsymbol{\mu}^{(\ell)} = \hat{\Pi}_\ell \mathbf{M}^{(\ell)}
,
\nonumber
\end{align}
obtained from the tensorial moments $\mathbf{M}^{(\ell)}$ using the trace-removal projector $\hat{\Pi}_\ell$,
and we examine in more depth their relation to the other objects in play.

The information contained in $\boldsymbol{\mu}^{(\ell)}$ is equivalent to that contained in the (`standard') spherical multipolar moments 
\begin{align}
M_{\ell m}
=
\int
S_{\ell m}(\vbr)
\rho(\vbr)
\d\vbr
\propto
\int
r^\ell
Y_{\ell m}(\theta,\phi)
\rho(\vbr)
\d\vbr
,
\end{align}
given by the integral of $\rho(\vbr)$ multiplied by the spherical harmonic $Y_{lm}(\theta,\phi)$.
For our purposes, the handling is much cleaner if we phrase this in terms of the solid harmonic
\begin{equation}
S_{\ell m}(\vbr)
=
\sqrt{
  \frac{4\pi}{2\ell+1}
  }
\:
r^\ell
\:
Y_{\ell m}(\theta,\phi)
,
\label{s-lm-definition}
\end{equation}
as the latter is a homogeneous polynomial of degree $\ell$ in the Cartesian components $(x,y,z)$ of $\vbr$.

As a homogeneous polynomial, the solid harmonic can be expressed explicitly as a sum of monomials,
\begin{align}
S_{\ell m}(\vbr)
=
\sum_{i_1,\ldots,i_\ell}
(s_{\ell m})_{i_1 \cdots i_\ell}
\;
x_{i_1} \cdots x_{i_\ell}
,
\end{align}
where the individual monomial coefficients $(s_{\ell m})_{i_1 \cdots i_\ell}$ can be assembled as the components of a tensor 
\begin{align}
\hat{\vb{s}}_{\ell m} = 
\sum_{i_1,\ldots,i_\ell}
(s_{lm})_{i_1 \cdots i_\ell}
\ue{i_1} \otimes \cdots \otimes \ue{i_\ell}
\end{align}
which returns $S_{\ell m}(\vbr)$ when contracted with $\vbr^{\otimes \ell}$:
\begin{align}
S_{\ell m}(\vbr)
& = 
\hat{\vb{s}}_{\ell m}
\tensordot
\vbr^{\otimes \ell}
.
\label{Slm-as-tlm-contraction}
\end{align}
Moreover, this tensor can be found directly from the explicit polynomial expressions for $S_{\ell m}(\vbr)$~\cite[\S5.1.7, though see also \S5.2.3]{Varshalovich1988}
\begin{align}
S_{\ell m}(\vbr)
& =
\sqrt{(l+m)!(l-m)!}
\nonumber \\ & \quad \times 
\sum_{p,q,r}
\frac{1}{p!q!r!}
\left(-\frac{x+iy}{2}\right)^p
\left(\frac{x-iy}{2}\right)^q
z^r
,
\end{align}
where the summation indices are restricted to $p+q+r = \ell$, $p-q = m$;
from this, we can simply infer directly the expression
\begin{align}
\hat{\vb{s}}_{\ell m}
& =
\sqrt{(\ell+m)!(\ell-m)!}
\sum_{p,q,r}
\frac{
  2^{-\frac{p+q}{2}}
  }{
  p!q!r!
  }
\sym\mathopen{}\left[
  \ue{+}^{\otimes p}
  {\otimes}
  \ue{-}^{\otimes q}
  {\otimes}
  \ue{0}^{\otimes r}
  \right]
.
\label{s-lm-basis-definition}
\end{align}
by replacing each Cartesian components $x_i$ of $\vbr$ by the corresponding basis vectors $\ue{i}$, which are then tensor-multiplied together,
where the spherical unit basis vectors are defined as $\ue{\pm} = \mp \frac{1}{\sqrt{2}}(\ue{x}\pm i\ue{y})$ and $\ue{0} = \ue{z}$.

That said, it is convenient to reformulate these tensors slightly, by adding a normalization constant and taking the complex conjugate.
Thus, we define
\begin{align}
\hat{\vb{t}}_{\ell m}
& =
\sqrt{\frac{\ell!}{(2\ell-1)!!}}
\,
\hat{\vb{s}}_{\ell m}^*
,
\label{t-lm-basis-definition}
\end{align}
which we term the $\hat{\vb{t}}_{\ell m}$ the multipolar basis tensors.
These obey the following basic properties:
\begin{subequations}%
\vspace{0.2em}%
\begin{itemize}[nosep,left=3pt]
\item 
They are, of course, purely multipolar, so
\begin{align}
\hat{\Pi}_\ell \hat{\vb{t}}_{\ell m}
=
\hat{\vb{t}}_{\ell m}
.
\label{t-lm-multipolarity}
\end{align}

\item
They are symmetric under complex conjugation via
\begin{align}
\hat{\vb{t}}_{\ell m}^*
=
(-1)^m
\hat{\vb{t}}_{\ell,-m}
.
\end{align}

\item
They are orthonormal,
\begin{align}
\hat{\vb{t}}_{\ell m}^* \tensordot \hat{\vb{t}}_{\ell m'}
=
\delta_{m,m'}
.
\end{align}

\item 
They satisfy the completeness relation
\begin{align}
\sum_m
\hat{\vb{t}}_{\ell m} 
(\hat{\vb{t}}_{\ell m}^* \tensordot \mathbf{A}^{(\ell)})
=
\hat{\Pi}_\ell \mathbf{A}^{(\ell)}
,
\label{t-lm-completeness}
\end{align}
where $\mathbf{A}^{(\ell)}$ is an arbitrary tensor of rank $\ell$,
with the sum
adding up to the $\ell$-polar projector $\hat{\Pi}_\ell$ of rank $\ell$.
\end{itemize}
\vspace{0.5em}

\end{subequations}

More importantly, if we integrate Eq.~\eqref{Slm-as-tlm-contraction} against $\rho(\vbr)$, we obtain a clean relation between the spherical multipolar moment $M_{\ell m}$ and the tensorial moment $\mathbf{M}^{(\ell)}$,
\begin{align}
M_{\ell m}
& = 
\sqrt{\frac{(2\ell-1)!!}{\ell!}}
\,
\hat{\vb{t}}_{\ell m}^*
\tensordot
\mathbf{M}^{(\ell)}
.
\label{M-lm-as-contraction-M}
\end{align}
Furthermore, using Eq.~\eqref{t-lm-multipolarity} and the fact that the projector $\hat{\Pi}_\ell$ is orthogonal (so $\hat{\Pi}_\ell\mathbf{A} \tensordot \mathbf B = \mathbf{A} \tensordot \hat{\Pi}_\ell\mathbf B$), we can rephrase \eqref{M-lm-as-contraction-M} as
\begin{align}
M_{\ell m}
& = 
\sqrt{\frac{(2\ell-1)!!}{\ell!}}
\,
\hat{\vb{t}}_{\ell m}^*
\tensordot
\boldsymbol{\mu}^{(\ell)}
.
\label{M-lm-as-contraction-mu}
\end{align}
Similarly, we can use the completeness relation \eqref{t-lm-completeness} to obtain
\begin{align}
\boldsymbol{\mu}^{(\ell)}
=
\sqrt{\frac{\ell!}{(2\ell-1)!!}}
\sum_m
M_{\ell m}
\hat{\vb{t}}_{\ell m} 
\label{mu-from-Mlm}
\end{align}
which fully encapsulates the claim that the spherical multipolar moments $M_{\ell m}$ are a minimal set of linearly-independent components of the tensorial multipolar moment~$\boldsymbol{\mu}^{(\ell)}$.

With this in hand, we can now examine our chirality measure, the traceless chiral moments from Eq.~\eqref{h-l1l2l3-definition}, which combine the tensorial multipolar moments into the pseudoscalar
$
h_{\ell_1 \ell_2 \ell_3}
=
\left( 
  \boldsymbol{\mu}^{(\ell_1)}
  \tensorcross
  \boldsymbol{\mu}^{(\ell_2)}
  \right)^{(\ell_3)}
\tensordot
\boldsymbol{\mu}^{(\ell_3)}
$
using our tensor triple product.
With the tooling we have just developed, we can use the basis expansion \eqref{mu-from-Mlm} to reduce $h_{\ell_1 \ell_2 \ell_3}$ to products of the spherical multipolar moments $M_{\ell m}$, which gives us
\begin{align}
h_{\ell_1 \ell_2 \ell_3}
& =
\sqrt{
  \frac{\ell_1!}{(2\ell_1-1)!!}
  \frac{\ell_2!}{(2\ell_2-1)!!}
  \frac{\ell_3!}{(2\ell_3-1)!!}
  }
\nonumber \\ & \qquad \times
\sum_{m_1, m_2, m_3}
M_{\ell_1 m_1}
M_{\ell_2 m_2}
M_{\ell_3 m_3}
\nonumber \\ & \qquad \qquad \times
\left( 
  \hat{\vb{t}}_{\ell_1 m_1}
  \tensorcross
  \hat{\vb{t}}_{\ell_2 m_2}
  \right)^{(\ell_3)}
\tensordot
\hat{\vb{t}}_{\ell_3 m_3}
.
\label{h-l1l2l3-to-tlm-cross-product}
\end{align}
This is a crucial relationship: 
it is structured as a superposition of cubic products of spherical multipolar moments, $M_{\ell m}$, which combine to form a (pseudo)scalar;
as mentioned in the main text, this has strong consequences.

Specifically,
the relationship~\eqref{h-l1l2l3-to-tlm-cross-product} is structured as a linear superposition of the components of the triple tensor product $M_{\ell_1 m_1}M_{\ell_2 m_2}M_{\ell_3 m_3}$,
where each of the tensor factors comes from an irreducible representation of the rotation group $\SO{3}$,
and where the combined sum is a scalar, i.e., is in the $\ell=0$ representation of the triple-tensor-product representation.
The representation theory of the rotational group~\cite{Hall2004} then enforces a uniqueness theorem, which requires the coefficients in the superposition to be multiples of the Wigner $3j$ symbols.
In other words, Eq.~\eqref{h-l1l2l3-to-tlm-cross-product} requires the identity
\begin{align}
\left( 
  \hat{\vb{t}}_{\ell_1 m_1}
  \tensorcross
  \hat{\vb{t}}_{\ell_2 m_2}
  \right)^{(\ell_3)}
\tensordot
\hat{\vb{t}}_{\ell_3 m_3}
& =
%
N_{\ell_1 \ell_2 \ell_3}
\begin{pmatrix}
  \ell_1 & \ell_2 & \ell_3 \\
  m_1 & m_2 & m_3
\end{pmatrix}
,
\label{tlm-cross-to-wigner-3j}
\end{align}
for some normalization constant $N_{\ell_1 \ell_2 \ell_3}$.%
\footnote{
We find the result~\eqref{tlm-cross-to-wigner-3j} to be remarkable as well as highly useful in forming and cementing intuition:
it illuminates the `true nature' of the Wigner $3j$ symbols, which are revealed as simply the cross product between the multipolar basis tensors;
and, separately,
it helps understand that cross product, as it is just the familiar Wigner $3j$ symbol.
(Of course, only one of those directions is likely to be useful, depending on whether one's intuitive understanding of the Wigner $3j$ symbols or of the tensor cross product is stronger.)
}

\subsection{Tensorial moments for polynomial distributions}
\label{app-tensorial-moments-poly-dist}
To wrap up our summary of the properties of multipolar tensors, in this section we turn to some useful properties that apply when the distribution itself is a well-behaved polynomial.

Thus, in particular, we consider distributions of the form
\begin{align}
\rho(\vbr) 
& = 
\tens{C}^{(k)}\tensordot \vbr^{\otimes k} f(r)
\nonumber \\ & =
C^{(k)}_{i_1\cdots i_k}
x_{i_1} \cdots x_{i_k}
f(r)
,
\end{align}
where the polynomial coefficients can be enclosed as a symmetric tensor $\tens{C}^{(k)}$ of rank $k$,
and we also include a sphe\-ri\-ca\-lly-\allowbreak{}symmetric function $f(r)$ 
(so that e.g.\ 
the distribution can be confined to a sphere, with $f(r) = \delta(r-1)$, 
or simply to ensure convergent integrals through, say, a gaussian filter $f(r) = e^{-r^2/\sigma^2}$).

We then ask the question: 
what are the tensorial moments $\tens{M}^{(n)} = \int \vbr^{\otimes n} \rho(\vbr)\d\vbr$ of $\rho(\vbr)$,
and how do they relate to our initial coefficient $\tens{C}^{(k)}$?
Standard intuition tells us that if $\tens{C}^{(k)}$ is strictly $\ell$-polar (and thus $\rho(\vbr)$ is proportional to a superposition of spherical harmonics with well-defined $\ell$), then the tensorial moment $\tens{M}^{(k)}$ must coincide with the $\tens{C}^{(k)}$, and this is indeed the case.
But what happens when $\tens{C}^{(k)}$ is not strictly multipolar?

The proof is straightforward but cumbersome, and we defer it to section~\ref{app-tensorial-moments-poly-dist-proof}.
The overall result rides on the standard intuition for fully-multipolar distributions: 
first decompose $\tens{C}^{(k)}$ into its multipolar components, and each of these will translate transparently to the corresponding moment tensor $\tens{M}^{(n)}$.

The most straightforward case is when $k=n$, in which case the relationship reads
%
\begin{align}
\tens{M}^{(n)}
& = 
\sum_{\ell}
\frac{
  4\pi \ell! \: 
  }{
  (2\ell+1)!!
  }
b_{\ell,\frac{n-\ell}{2}} 
\ 
\projj{n}{\ell}
\tens{C}^{(n)}
\nonumber \\ & \quad \times
\int_0^\infty
r^{2n+2}
f(r)
\d r
,
\label{moment-from-poly-distro-k-eq-n}
\end{align}
i.e., $\tens{M}^{(n)}$ is a superposition of the $\ell$-polar components of $\tens{C}^{(k)}$, with $\ell$-dependent coefficients, 
multiplied by the radial integral
$
\int_0^\infty
r^{2n+2}
f(r)
\d r
$.
Of course, as promised, when $\tens{C}^{(\ell)} = \projj{\ell}{\ell}\tens{C}^{(\ell)}$ has well-defined multipolarity, this result simplifies further, to
\begin{align}
\tens{M}^{(\ell)}
& = 
\frac{
  4\pi \ell! \: 
  }{
  (2\ell+1)!!
  }
\int_0^\infty \!\!\!\!
r^{2\ell+2}
f(r)
\d r
\ 
\tens{C}^{(\ell)}
.
\end{align}

Things are more complicated in the case when $k\neq n$, as then the $\tens{C}^{(k)}$ need to be traced down, or lifted up, to match the rank $n$ of the desired moment.
Thus, for the general case, as we prove in section~\ref{app-tensorial-moments-poly-dist-proof}, we have
\begin{align}
\tens{M}^{(n)}
& = 
\sum_{\ell}^{\min(n,k)}
\frac{4\pi \ell!}{(2\ell+1)!!}
b_{\ell,\frac{n-\ell}{2}} 
b_{\ell,\frac{k-\ell}{2}} 
\int_0^\infty
r^{n+k+2}
f(r)
\d r
\nonumber \\ & \quad \times
\lift^{\frac{n-\ell}{2}}(
  \projj{\ell}{\ell}
  \Tr^{\frac{k-\ell}{2}}(
    \tens{C}^{(k)}
    )
  )
.
\label{moment-from-poly-distro-general}
\end{align}
The lift and the trace operators that appear here are partial inverses, but they appear to different orders.
Thus, this can simplify if we restrict it to the cases when $n>k$, where we have
%
%
%
%
\begin{align}
\tens{M}^{(n)}
& = 
\sum_{\ell}
\frac{4\pi \ell!}{(2\ell+1)!!}
b_{\ell,\frac{n-\ell}{2}} 
\:
\lift^{\frac{n-k}{2}}(
  \projj{k}{\ell}
  \tens{C}^{(k)}
  )
\nonumber \\ & \quad \times
\int_0^\infty
r^{n+k+2}
f(r)
\d r
.
\label{moment-from-poly-distro-k-lt-n}
\end{align}
%
and, similarly, for $n<k$,
\begin{align}
\tens{M}^{(n)}
& = 
\sum_{\ell}
\frac{4\pi \ell!}{(2\ell+1)!!}
b_{\ell,\frac{k-\ell}{2}} 
\:
\projj{n}{\ell}
\Tr^{\frac{k-n}{2}}(
  \tens{C}^{(k)}
  )
\nonumber \\ & \quad \times
\int_0^\infty
r^{n+k+2}
f(r)
\d r
.
\label{moment-from-poly-distro-k-gt-n}
\end{align}

\subsection{Tensorial moments for polynomial distributions: rigorous proof}
\label{app-tensorial-moments-poly-dist-proof}

As in section~\ref{app-tensorial-moments-poly-dist}, in this section we onsider a distribution of the form 
$
\rho(\vbr) = \tens{C}^{(k)}\tensordot \vbr^{\otimes k} f(r)
$
for some symmetric tensor $\tens{C}^{(k)}$ of rank $k$ and a sphe\-ri\-ca\-lly-\allowbreak{}symmetric function $f(r)$,
and aim to calculate its rank-$n$ tensorial moment $\tens{M}^{(n)} = \int \vbr^{\otimes n} \rho(\vbr)\d\vbr$.

To do this, we expand both $\tens{M}^{(n)}$ and $\tens{C}^{(k)}$ as superpositions of their multipolar components, 
after which we can 
shift the multipolar projector to a lower tensor rank using~\eqref{projj-as-trace-proj-lift},
\begin{align}
\tens{M}^{(n)}
& = 
\int 
\vbr^{\otimes n} 
\rho(\vbr)
\d\vbr
\nonumber \\ & = 
\sum_{\ell,\ell'}
\int 
\left(
  \projj{n}{\ell}
  \vbr^{\otimes n} 
  \right)
\left(
  \projj{k}{\ell'}
  \tens{C}^{(k)}
  \right)
\tensordot \vbr^{\otimes k} 
f(r)
\d\vbr
\nonumber \\ & = 
\sum_{\ell,\ell'}
\int 
\left(
  b_{\ell,\frac{n-\ell}{2}} 
  \lift^{\frac{n-\ell}{2}}(
    \projj{\ell}{\ell}
    \Tr^{\frac{n-\ell}{2}}(
      \vbr^{\otimes n}
      )
    )
  \right)
\nonumber \\ & \quad {\times}
\left(
  b_{\ell',\frac{k-\ell'}{2}} 
  \lift^{\frac{k-\ell'}{2}}(
    \projj{\ell'}{\ell'}
    \Tr^{\frac{k-\ell'}{2}}(
      \tens{C}^{(k)}
      )
    )
  \right)
\tensordot \vbr^{\otimes k} 
\!
f(r)
\d\vbr
\nonumber \\ & = 
\sum_{\ell,\ell'}
b_{\ell,\frac{n-\ell}{2}} 
b_{\ell',\frac{k-\ell'}{2}} 
\lift^{\frac{n-\ell}{2}}
\int \!
\d\vbr \:
f(r)
r^{k-\ell'}
r^{n-\ell}
\projj{\ell}{\ell}
\vbr^{\otimes \ell}
\nonumber \\ & \quad \times
\left(
  \projj{\ell'}{\ell'}
  \Tr^{\frac{k-\ell'}{2}}(
    \tens{C}^{(k)}
    )
  \right)
\tensordot \vbr^{\otimes k} 
\end{align}
and then use the completeness relation~\eqref{t-lm-completeness} for the multipolar projectors:
\begin{align}
\tens{M}^{(n)}
& = 
\sum_{\ell,\ell'}
\sum_{m,m'}
b_{\ell,\frac{n-\ell}{2}} 
b_{\ell',\frac{k-\ell'}{2}} 
\lift^{\frac{n-\ell}{2}}(
  \hat{\vb{t}}_{\ell m} 
  )
\nonumber 
\\ & \quad \times
\left(
  \hat{\vb{t}}_{\ell' m'}^* 
  \tensordot 
  \Tr^{\frac{k-\ell'}{2}}(
    \tens{C}^{(k)}
    )
  \right)
\\ & \quad \times
\nonumber 
\int 
(\hat{\vb{t}}_{\ell m}^* \tensordot \vbr^{\otimes \ell})
\left(
  \hat{\vb{t}}_{\ell' m'} 
  \tensordot 
  \vbr^{\otimes k} 
  \right)
r^{k-\ell'}
r^{n-\ell}
f(r)
\d\vbr 
,
\end{align}
which then becomes, using~\eqref{t-lm-basis-definition}, \eqref{Slm-as-tlm-contraction} and~\eqref{s-lm-definition},
\begin{align}
\tens{M}^{(n)}
& = 
%
\sum_{\ell,\ell'}
\sum_{m,m'}
\sqrt{
  \frac{4\pi \ell!}{(2\ell+1)!!}
  \frac{4\pi \ell'!}{(2\ell'+1)!!}
  }
b_{\ell,\frac{n-\ell}{2}} 
b_{\ell',\frac{k-\ell'}{2}} 
\nonumber 
\\ & \quad \times
\lift^{\frac{n-\ell}{2}}(
  \hat{\vb{t}}_{\ell m} 
  )
\left(
  \hat{\vb{t}}_{\ell' m'}^* 
  \tensordot 
  \Tr^{\frac{k-\ell'}{2}}(
    \tens{C}^{(k)}
    )
  \right)
\\ & \quad \times
\nonumber 
\int 
Y_{\ell m}(\vbr)
Y_{\ell' m'}^*(\vbr)
r^{n+k}
f(r)
\d\vbr 
,
\end{align}
and, since
$
\int 
Y_{\ell m}(\vbr)
Y_{\ell' m'}^*(\vbr)
r^{n+k}
f(r)
\d\vbr 
$
is diagonal and equal to
$
\delta_{\ell \ell'}
\delta_{m m'}
\int_0^\infty
r^{n+k+2}
f(r)
\d r
$,
\begin{align}
\tens{M}^{(n)}
& = 
\sum_{\ell,m}
\frac{4\pi \ell!}{(2\ell+1)!!}
b_{\ell,\frac{n-\ell}{2}} 
b_{\ell,\frac{k-\ell}{2}} 
\int_0^\infty
r^{n+k+2}
f(r)
\d r
\nonumber 
\\ & \quad \times
\lift^{\frac{n-\ell}{2}}(
  \hat{\vb{t}}_{\ell m} 
  \left(
    \hat{\vb{t}}_{\ell m}^* 
    \tensordot 
    \Tr^{\frac{k-\ell}{2}}(
      \tens{C}^{(k)}
      )
    \right)
  )
,
\end{align}
where the limits of the $\ell,\ell'$ summations transform via $\sum_{\ell}^n\sum_{\ell'}^k \delta_{\ell \ell'} = \sum_{\ell}^{\min(n,k)}$.

Finally, we can remove the $m$ dependence by re-applying the completeness relation~\eqref{t-lm-completeness}, and this gives us the full result~\eqref{moment-from-poly-distro-general} from the overview section.

\subsection{The trace-removal projector: rigorous proofs}
\label{app-trace-removal-projector-proofs}
In this section, we provide a fuller proof of the values of the coefficients $c_{n,\ell}$ and $b_{\ell,m}$, starting with the general identity~\eqref{trace-of-sym-basic}.
The simplest place to start is the representation with maximal angular momentum number $\ell = n$, which is defined (within this framework) as the `simple' case, where $\Tr(\tens{A}) = 0$ vanishes.
For that case, the identity~\eqref{trace-of-sym-basic} is already sufficient to determine the coefficient of interest,
\begin{align}
c_{\ell,\ell}
=
\frac{
  (\ell+2)(\ell+1)
  }{
  2(2\ell+d)
  }
,
\end{align}
as the second term vanishes.

For representations with higher rank than angular momentum (i.e.\ $n>\ell$), as indicated in Fig.~\ref{fig:lifts-and-traces}, we work recursively:
we iterate the identity~\eqref{trace-of-sym-basic} until we reach tensors of rank $\ell$, where the trace will vanish, and use this to compute the chain of coefficients.
Thus, we rephrase Eq.~\eqref{trace-of-sym-basic} into the form
\begin{align}
\Tr[\lift(\tens{A})]
 =
u_n
\tens{A}
& 
+
v_n
\lift(\Tr(\tens{A}))
\label{trace-of-sym-elegant}
\\ \nonumber
\text{with}
\quad
u_n
=
\frac{
  2(2n+d)
  }{
  (n+2)(n+1)
  }
,
&
\ 
v_n
=
\frac{
  n(n-1)
  }{
  (n+2)(n+1)
  }
,
\end{align}
which we can then apply recursively.
Thus, as the first step, by applying this form but replacing $\tens{A}$ with $\lift(\tens{A})$, we get
\begin{align}
\Tr[\lift^2(\tens{A})]
& =
u_{n+2}
\lift(\tens{A})
+
v_{n+2}
\lift(\Tr(\lift(\tens{A})))
\nonumber \\ & =
u_{n+2}
\lift(\tens{A})
+
v_{n+2}
\lift(
  u_n
  \tens{A}
  +
  v_n
  \lift(\Tr(\tens{A}))
  )
\nonumber \\ & =
(u_{n+2} {+} v_{n+2}u_n)
\lift(\tens{A})
+
v_{n+2}
v_n
\lift^2(\Tr(\tens{A}))
.
\label{trace-lift-iteration-1}
\end{align}
for $\tens{A}\in \symtens{n}$ of rank $n$.

To apply this, we set $\tens{A} \in \irrep{\ell}{\ell}$, giving us a tensor $\lift(\tens{A}) \in \irrep{\ell+2}{\ell}$
with the property that $\Tr(\tens{A})=0$, 
so that we can read off the next coefficient of interest, 
\begin{align}
c_{\ell+2,\ell}
& =
(u_{\ell+2} + v_{\ell+2}u_\ell)^{-1}
\nonumber \\ & =
%
\frac{
  (\ell+4)(\ell+3)
  }{
  2^2(2\ell+2+d)
  }
,
\end{align}
which determines $c_{\ell+2,\ell}\Tr$ as the inverse of $\lift$ on $\irrep{\ell+2}{\ell}$.
Moreover, we can also get the inverse of the \textit{repeated} lift operator, $\lift^2$, on $\irrep{\ell}{\ell}$, given by $b_{\ell,2}\Tr^2$, with coefficient
\begin{align}
b_{\ell,2}
& =
c_{\ell+2,2}
c_{\ell,2}
\nonumber \\ & =
\frac{
  (\ell+4)(\ell+3)
  }{
  2^2(2\ell+2+d)
  }
\frac{
  (\ell+2)(\ell+1)
  }{
  2(2\ell+d)
  }
\nonumber \\ & 
=
\frac{
  (\ell+4)!
  }{
  2^3
  \ell!
  }
\frac{
  (2\ell-2+d)!!
  }{
  (2\ell+2+d)!!
  }
.
\end{align}
A similar argument then proves the general case by induction.

\section{Perturbation theory derivation}
\label{sec-appendix-perturbation-theory}
As described in Section~\ref{sec-photoelectron-spectra}, here we consider in detail the perturbation-theory calculation for chiral photoionization.
We use the simplest example of photoionization driven by a chiral field: resonantly-enhanced two-photon ionization of atomic hydrogen driven by an elliptically-polarized fundamental, combined with a second harmonic with linear polarization orthogonal to the fundamental's plane of ellipticity.

Thus, we consider a hydrogen atom, initially in its ground state, ionized by the bichromatic $\omega:2\omega$ field
\begin{align}
\vbE(t)
=
\Re\mathopen{}\left[
\vbE_1 e^{-i\omega t} 
+
\vbE_2 e^{-2i\omega t} 
\right]
f(t),
\end{align}
where $\vbE_1$ and $\vbE_2$ are complex field-polarization amplitudes, 
and $f(t) = \exp(-t^2/2T^2)$ is a gaussian envelope.
We tune the fields close to the $1s$-$2p$ transition, with a detuning $\delta\omega = \omega - (E_{2p}-E_{1s})/\hbar$.

In these conditions, the dynamics will be confined to the bound $2p$ manifold, and to the ionized continuum including s, p and d waves (with $\ell=0,1$ and 2, respectively), 
centred at energy $\frac{1}{2m_e}p_0^2 = E_{1s} + 2\omega$,
which we write in the form
\begin{align}
\ket{\psi(t)}
& =
a_{100}(t) 
e^{-i E_{1s}t/\hbar} 
\ket{100}
\nonumber \\ & \qquad
+\sum_m 
a_{21m}(t) 
e^{-i E_{2p}t/\hbar} 
\ket{21m}
\nonumber \\ & \qquad
+ 
\sum_{\ell,m} 
\int_0^\infty \!\! \d p \ 
b_{\ell m}(p,t) 
e^{-\frac{ip^2t}{2m_e\hbar}}
\ket{p,\ell m}
.
\end{align}
Our observable of interest is the momentum-resolved wavefunction,
$\psi(\vbp) = \braket{\vbp}{\psi(t\to\infty)}$,
evaluated at the end of the pulse, 
and its corresponding population density,
$\rho(\vbp) = |\psi(\vbp)|^2$.

In the rotating-wave approximation, the hamiltonian of the system,
$\hat{H} = \hat{H}_0 + q_e\hat{\vb{r}}\cdot \vbE(t)$,
induces 
the bound-bound transition from $1s$ to $2p$,
from $1s$ to the $p$-wave continuum,
and from $2p$ to the $s$- and $d$-wave continua.
Thus, the Schrödinger equation reads
\begin{widetext}
\begin{align}
i\hbar \, \dot a_{21m}(t)
& = 
\frac12
e^{-i\delta\omega \, t}
f(t)
\matrixel{21m}{q_e\hat{\vb{r}}}{100}
\cdot 
\vbE_1 
a_{100}(t)
\\ 
i\hbar \, \dot b_{00}(p,t)
& =
\frac12
e^{-i \frac{p^2 - p_0^2}{2m_e\hbar} t}
f(t)
\sum_m
\matrixel{p,00}{q_e\hat{\vb{r}}}{21m}
\cdot 
\vbE_1
\, a_{21m}(t)
\\ 
i\hbar \, \dot b_{1m}(p,t)
& =
\frac12
e^{-i \frac{p^2 - p_0^2}{2m_e\hbar} t}
f(t)
\matrixel{p,1m}{q_e\hat{\vb{r}}}{100}
\cdot 
\vbE_2
a_{100}(t)
\\ 
i\hbar \, \dot b_{2m}(p,t)
& =
\frac12
e^{-i \frac{p^2 - p_0^2}{2m_e\hbar} t}
f(t)
\sum_{m'}
\matrixel{p,2m}{q_e\hat{\vb{r}}}{21m'}
\cdot 
\vbE_1
\, a_{21m'}(t)
,
\end{align}
We solve this system within perturbation theory, assuming that $a_{100}(t) \equiv 1$, from which we can then read directly the solutions
{\allowdisplaybreaks%
\begin{align}
a_{21m}(t)
& = 
\frac{1}{2i\hbar}
\int_{-\infty}^{t} \!\! \d t'
e^{-i\delta\omega \, t'}
f(t')
\matrixel{21m}{q_e\hat{\vb{r}}}{100}
\cdot 
\vbE_1 
\\
b_{1m}(p,t)
& =
\frac{1}{2i\hbar}
\int_{-\infty}^{t} \!\! \d t'
e^{-i \frac{p^2 - p_0^2}{2m_e\hbar} t'}
f(t')
\matrixel{p,1m}{q_e\hat{\vb{r}}}{100}
\cdot 
\vbE_2
\label{one-photon-components-intermediate}
\end{align}
}%
for the $p$-state components,
and from these we obtain the $s$- and $d$-state components as
{\allowdisplaybreaks%
\begin{align}
b_{00}(p,t)
& =
-\frac{1}{4\hbar^2}
\int_{-\infty}^{t} \!\! \d t'
e^{-i \frac{p^2 - p_0^2}{2m_e\hbar} t'}
f(t')
\sum_m
\matrixel{p,00}{q_e\hat{\vb{r}}}{21m}
\cdot 
\vbE_1
\int_{-\infty}^{t'} \!\! \d t''
e^{-i\delta\omega \, t''}
f(t'')
\matrixel{21m}{q_e\hat{\vb{r}}}{100}
\cdot 
\vbE_1 
\nonumber \\ & =
-\frac{1}{4\hbar^2}
\left(
\sum_m
\matrixel{p,00}{q_e\hat{\vb{r}}}{21m}
\cdot 
\vbE_1
\matrixel{21m}{q_e\hat{\vb{r}}}{100}
\cdot 
\vbE_1 
\right)
\left(
\int_{-\infty}^{t} \!\! \d t'
e^{-i \frac{p^2 - p_0^2}{2m_e\hbar} t'}
f(t')
\int_{-\infty}^{t'} \!\! \d t''
e^{-i\delta\omega \, t''}
f(t'')
\right)
\\
b_{2m}(p,t)
& =
-\frac{1}{4\hbar^2}
\int_{-\infty}^{t} \!\! \d t'
e^{-i \frac{p^2 - p_0^2}{2m_e\hbar} t'}
f(t')
\sum_{m'}
\matrixel{p,2m}{q_e\hat{\vb{r}}}{21m'}
\cdot 
\vbE_1
\int_{-\infty}^{t'} \!\! \d t''
e^{-i\delta\omega \, t''}
f(t'')
\matrixel{21m'}{q_e\hat{\vb{r}}}{100}
\cdot 
\vbE_1 
\nonumber \\ & =
-\frac{1}{4\hbar^2}
\left(
\sum_{m'}
\matrixel{p,2m}{q_e\hat{\vb{r}}}{21m'}
\cdot 
\vbE_1
\matrixel{21m'}{q_e\hat{\vb{r}}}{100}
\cdot 
\vbE_1 
\right)
\left(
\int_{-\infty}^{t} \!\! \d t'
e^{-i \frac{p^2 - p_0^2}{2m_e\hbar} t'}
f(t')
\int_{-\infty}^{t'} \!\! \d t''
e^{-i\delta\omega \, t''}
f(t'')
\right)
.
\label{two-photon-components-intermediate}
\end{align}
}%
Both of these components are cleanly separated into two factors~-- one encoding the directional dependence, and the second containing the temporal and energy information.

These form a complete solution to the problem, and allow us to reconstruct the final wavefunction,
using the partial-wave expansion
\begin{align}
\ket{\vb p}
=
\sum_{lm}
Y_{lm}(\hat{\vb p})^*
\ket{p,lm}
\end{align}
for the scattering states, giving
\begin{align}
\psi(\vbp)
& =
\sum_{\ell,m}
Y_{lm}(\hat{\vbp})
b_{\ell m}(p,\infty) 
.
\end{align}
The dependence on the pulse parameters is fully contained within two functions of the photoelectron momentum $p$, which correspond to the temporal factors in~\eqref{one-photon-components-intermediate} and~\eqref{two-photon-components-intermediate},
\begin{subequations}
\begin{align}
c_1(p)
& = 
\frac{1}{2i\hbar}
\int_{-\infty}^{\infty} \!\! \d t'
e^{-i\delta\omega \, t'}
f(t')
\\
c_2(p)
& =
-\frac{1}{4\hbar^2}
\int_{-\infty}^{\infty} \!\! \d t'
e^{-i \frac{p^2 - p_0^2}{2m_e\hbar} t'}
f(t')
\int_{-\infty}^{t'} \!\! \d t''
e^{-i\delta\omega \, t''}
f(t'')
\end{align}
\label{PT-c1-c2-def}
\end{subequations}
encoding the one- and two-photon processes, respectively.
This allows us to write the final wavefunction in the form
\begin{align}
\psi(\vbp)
& =
c_1(p)
\sum_{m}
Y_{1m}(\hat{\vbp})
\matrixel{p,1m}{q_e\hat{\vb{r}}}{100}
\cdot 
\vbE_2
\nonumber \\ & \quad +
c_2(p)
\sum_m
Y_{00}(\hat{\vbp})
\matrixel{p,00}{q_e\hat{\vb{r}}}{21m}
\cdot 
\vbE_1
\matrixel{21m}{q_e\hat{\vb{r}}}{100}
\cdot 
\vbE_1 
\nonumber \\ & \quad +
c_2(p)
\sum_{m,m'}
Y_{2m}(\hat{\vbp})
\matrixel{p,2m}{q_e\hat{\vb{r}}}{21m'}
\cdot 
\vbE_1
\matrixel{21m'}{q_e\hat{\vb{r}}}{100}
\cdot 
\vbE_1 
.
\label{pt-final-state-intermediate}
\end{align}

Our main focus here is on the angular dependence, which is encoded in the dipole transition matrix elements. 
To deal with these cleanly, we use the spherical basis vectors, 
$\ue{\pm}=\mp \tfrac{1}{\sqrt{2}}(\ue x \pm i \ue y)$ and $\ue0=\ue z$,
with which we can write any arbitrary vector $\vb v$ in the form
\begin{equation}
\vb v
=
\sum_{q=-1}^1
v_q
\ue{q}^*
=
\sqrt{\frac{4\pi}{3}}
v
\sum_{q=-1}^1
Y_{1q}(\hat{\vb v})
\ue{q}^*
,
\end{equation}
and dot products between real-valued vectors as
$\vb u \cdot \vb v = \sum_{q=-1}^1 u_q^* v_{q} = \sum_{q=-1}^1 u_q v_{q}^*$%
.
This allows us to write all of the matrix elements, via the Wigner-Eckart theorem~\cite[\S13.1.1]{Varshalovich1988}, in the form
{\allowdisplaybreaks%
\begin{align}
\matrixel{p,1m}{\hat{\vb{r}}}{100}
\cdot 
\vbE_2
& =
\sqrt{\frac{4\pi}{3}}
\sum_{q=-1}^1
\matrixel{p,1m}{ \hat r Y_{1q}(\hat{\vb r}) }{100}
\ue{q}^*
\cdot 
\vbE_2
=
\sqrt{\frac{4\pi}{3}}
\sum_{q=-1}^1
(-1)^{1-m}
\matrixel{p,1}{|\hat r|}{10}
\begin{pmatrix}
1 & 1 & 0 \\
-m & q & 0
\end{pmatrix}
\ue{q}^*
\cdot 
\vbE_2
\\
\matrixel{21m}{\hat{\vb{r}}}{100}
\cdot 
\vbE_1 
& =
\sqrt{\frac{4\pi}{3}}
\sum_{q=-1}^1
\matrixel{21m}{\hat r Y_{1q}(\hat{\vb r})}{100}
\ue{q}^*
\cdot 
\vbE_1 
=
\sqrt{\frac{4\pi}{3}}
\sum_{q=-1}^1
(-1)^{1-m}
\matrixel{21}{|\hat r|}{10}
\begin{pmatrix}
1 & 1 & 0 \\
-m & q & 0
\end{pmatrix}
\ue{q}^*
\cdot 
\vbE_1 
\\
\matrixel{p,00}{\hat{\vb{r}}}{21m}
\cdot 
\vbE_1
& =
\sqrt{\frac{4\pi}{3}}
\sum_{q=-1}^1
\matrixel{p,00}{ \hat r Y_{1q}(\hat{\vb r})}{21m}
\ue{q}^*
\cdot 
\vbE_1
=
\sqrt{\frac{4\pi}{3}}
\sum_{q=-1}^1
\matrixel{p,0}{|\hat r|}{21}
\begin{pmatrix}
0 & 1 & 1 \\
0 & q & m
\end{pmatrix}
\ue{q}^*
\cdot 
\vbE_1
\\
\matrixel{p,2m}{\hat{\vb{r}}}{21m'}
\cdot 
\vbE_1
& =
\sqrt{\frac{4\pi}{3}}
\sum_{q=-1}^1
\matrixel{p,2m}{ \hat r Y_{1q}(\hat{\vb r}) }{21m'}
\ue{q}^*
\cdot 
\vbE_1
=
\sqrt{\frac{4\pi}{3}}
\sum_{q=-1}^1
(-1)^{m}
\matrixel{p,2}{|\hat r|}{21}
\begin{pmatrix}
2 & 1 & 1 \\
-m & q & m'
\end{pmatrix}
\ue{q}^*
\cdot 
\vbE_1
.
\end{align}%
}%
And, finally, we can use these to write the combinations that appear in our final-state wavefunction~\eqref{pt-final-state-intermediate} in the form
{\allowdisplaybreaks%
\begin{align}
\sum_{m}
Y_{1m}(\hat{\vbp})
\matrixel{p,1m}{q_e\hat{\vb{r}}}{100}
\cdot 
\vbE_2
& =
\sqrt{\frac{4\pi}{3}}
\matrixel{p,1}{|q_e\hat r|}{10}
\sum_{m,q}
(-1)^{1-m}
Y_{1m}(\hat{\vbp})
\begin{pmatrix}
1 & 1 & 0 \\
-m & q & 0
\end{pmatrix}
\ue{q}^*
\cdot 
\vbE_2
%
%
\nonumber \\ & =
\frac{1}{\sqrt{3}p}
\matrixel{p,1}{|q_e\hat r|}{10}
\vbp
\cdot 
\vbE_2
\\
\sum_m
Y_{00}(\hat{\vbp})
\matrixel{p,00}{q_e\hat{\vb{r}}}{21m}
\cdot 
\vbE_1
\matrixel{21m}{q_e\hat{\vb{r}}}{100}
\cdot 
\vbE_1
& =
%
\frac{\sqrt{4\pi}}{3}
\matrixel{p,0}{|q_e\hat r|}{21}
\matrixel{21}{|q_e\hat r|}{10}
\nonumber \\ & \qquad \times
\sum_{m,q,q'}
(-1)^{1-m}
\begin{pmatrix}
0 & 1 & 1 \\
0 & q' & m
\end{pmatrix}
\begin{pmatrix}
1 & 1 & 0 \\
-m & q & 0
\end{pmatrix}
\ue{q'}^*
\cdot 
\vbE_1
\ue{q}^*
\cdot 
\vbE_1 
%
%
\nonumber \\ & =
-
\frac{\sqrt{4\pi}}{9}
\matrixel{p,0}{|q_e\hat r|}{21}
\matrixel{21}{|q_e\hat r|}{10}
\vbE_1
\cdot 
\vbE_1 
\\
\sum_{m,m'}
Y_{2m}(\hat{\vbp})
\matrixel{p,2m}{q_e\hat{\vb{r}}}{21m'}
\cdot 
\vbE_1
\matrixel{21m'}{q_e\hat{\vb{r}}}{100}
\cdot 
\vbE_1 
& =
\frac{4\pi}{3}
\matrixel{p,2}{|\hat r|}{21}
\matrixel{21}{|\hat r|}{10}
\sum_{m,m',q,q'}
(-1)^{1-m'-m}
\nonumber \\ & \qquad \times
Y_{2m}(\hat{\vbp})
\begin{pmatrix}
2 & 1 & 1 \\
-m & q' & m'
\end{pmatrix}
\begin{pmatrix}
1 & 1 & 0 \\
-m' & q & 0
\end{pmatrix}
\ue{q}^*
\cdot 
\vbE_1 
\ue{q'}^*
\cdot 
\vbE_1
%
\nonumber \\ & =
\frac{\sqrt{20\pi}}{3}
\frac{
  \matrixel{p,2}{|q_e \hat r|}{21}
  \matrixel{21}{|q_e \hat r|}{10}
  }{
  p^2
  }
\proj{2}\vbp^{\otimes 2}
\tensordot
\vbE_1^{\otimes 2}
.
\end{align}
}

This wraps up our calculations of the wavefunction, as these expressions can then be inserted directly into \eqref{pt-final-state-intermediate} to give us
%
%
%
\begin{align}
\psi(\vbp)
& =
c_\mathrm{p}(p)
\:
\vbp
\cdot 
\vbE_2
+
c_\mathrm{s}(p)
\:
\vbE_1
\cdot 
\vbE_1 
+
c_\mathrm{d}(p)
\:
\proj{2}\vbp^{\otimes 2}
\tensordot
\vbE_1^{\otimes 2}
\label{pt-final-wavefunction-final}
\end{align}
in terms of the spherically-symmetric amplitudes
\begin{subequations}
\begin{align}
c_\mathrm{p}(p)
& =
c_1(p)
\frac{1}{\sqrt{3}p}
\matrixel{p,1}{|q_e\hat r|}{10}
\\
c_\mathrm{s}(p)
& =
-
c_2(p)
\frac{\sqrt{4\pi}}{9}
\matrixel{p,0}{|q_e\hat r|}{21}
\matrixel{21}{|q_e\hat r|}{10}
\\
c_\mathrm{d}(p)
& =
c_2(p)
\frac{\sqrt{20\pi}}{3}
\frac{
  \matrixel{p,2}{|q_e \hat r|}{21}
  \matrixel{21}{|q_e \hat r|}{10}
  }{
  p^2
  }
.
\end{align}
\label{PT-cs-cp-cd-final}
\end{subequations}

To relate this to experiment, we now need to calculate the experimental observable, namely, the measured photoelectron momentum distribution $\rho(\vbp) = |\psi(\vbp)|^2$,
which reads
\begin{align}
\rho(\vbp)
& =
%
%
%
%
\left|
c_\mathrm{s}(p)
\:
\vbE_1
\cdot 
\vbE_1
\right|^2
+
\Re\mathopen{}
\left[
2
c_\mathrm{s}^*(p)
c_\mathrm{p}(p)
(
\vbE_1^*
\cdot 
\vbE_1^*
)
\vbE_2
\right]
\cdot 
\vbp
\nonumber \\ & \quad
+
\Re\mathopen{}
\left[
2
c_\mathrm{s}^*(p)
c_\mathrm{d}(p)
(
\vbE_1^*
\cdot 
\vbE_1^*
)
\proj{2}\vbE_1^{\otimes 2}
%
+
|c_\mathrm{p}(p)|^2
\vbE_2^* \otimes \vbE_2
\right]
\tensordot
\vbp^{\otimes 2}
\nonumber \\ & \quad
+
\Re\mathopen{}
\left[
2
c_\mathrm{p}^*(p)
c_\mathrm{d}(p)
\:
\sym(
  \vbE_2^*
  \otimes
  \proj{2}\vbE_1^{\otimes 2}
  )
\right]
\tensordot
\vbp^{\otimes 3}
\nonumber \\ & \quad
+
\Re\mathopen{}
\left[
|c_\mathrm{d}(p)|^2
\:
\proj{2}\vbE_1^{*\otimes 2}
\otimes
\proj{2}\vbE_1^{\otimes 2}
\right]
\tensordot
\vbp^{\otimes 4}
,
\label{rho-PT-final-result}
\end{align}
organized by the tensor rank of the power of $\vbp$ involved.

\end{widetext}

For generic polarization vectors $\vbE_1$ and $\vbE_2$, it is fully possible to extract 
the unabridged tensorial moments $\tens{M}^{(n)}$ 
(as well as the traceless tensorial multipolar moments $\boldsymbol{\mu}^{(\ell)}$)
for the distribution~\eqref{rho-PT-final-result}, 
following the definitions given in the main text and the formalism from Appendix~\ref{app-tensorial-moments-poly-dist}.
From the distribution~\eqref{rho-PT-final-result}, one can read off directly that 
the rank-4 moment~$\tens{M}^{(4)}$ will be a fourth-degree polynomial in $\vbE_1$ and $\vbE_1^*$
and the rank-3 moment~$\tens{M}^{(3)}$ will be a fourth-degree polynomial in $\vbE_1$, $\vbE_2$, and their conjugates,
coming from the coefficients in front of $\vbp^{\otimes 4}$ and $\vbp^{\otimes 3}$, respectively.
The rank-2 tensor~$\tens{M}^{(2)}$, coming from the $\vbp^{\otimes 2}$ term, contains polynomials of degrees 2 and 4 in these amplitudes, depending on the ionization channel.

These nonzero moments guide directly to $\chi_{234}$ as a natural chiral measure,%
\footnote{
The traceless moment $h_{234}$ also appears as a natural candidate, with identical considerations.
}
which, as a product of those three moments, must therefore emerge as a polynomial of mixed rank, containing both ninth- and eleventh-degree terms~-- in quite a combinatoric variety.
Those polynomials, while hard to instantly visualize for the general case, are an essential result, as they show a direct connection between the photoelectron momentum distribution 
and 
the chiral nonlinear correlation functions of the electromagnetic field which have been proposed~\cite{Ayuso2019} to quantify the chirality of synthetic chiral light;
in the notation of Ref.~\cite{Ayuso2019} they would have the form $h^{(9)}$ and $h^{(11)}$.

The field configuration we use is the same as was proposed previously for giant enantiosensitivity in high-order harmonic generation~\cite{Ayuso2019}, where the chirality of the field is determined by the \textit{fifth}-order chiral correlation function~($h^{(5)}$).
In our case, our results show that for ionization, in contrast, even the simplest possible configurations already involve, at least, chiral correlation functions of the ninth order and higher.
The reason is clear: in our formalism the experimental observable is the population density, which is a higher-order object, and there is no direct access to the wavefunction itself, or its coherence.

$\ $

These general considerations aside, we now turn to get a concrete idea of the behaviour of our chiral measures for the final distribution~\eqref{rho-PT-final-result}.
Even for the particular field configuration at play 
($\vbE_1 = E_1 \frac{1}{\sqrt{1+\epsilon^2}}(\ue{x} + i \epsilon \ue{y})$ 
elliptically polarized in the $xy$ plane,
and
$\vbE_2 = E_2 e^{i\varphi} \ue{z}$
orthogonal to that plane)
the tensorial moments $\tens{M}^{(n)}$ are rather cumbersome, but they can be handled easily through computer algebra; 
our implementation of that process is available as Ref.~\cite{figureMaker}.
The end result for the chiral moment $\chi_{234}$ is, then,
\begin{widetext}%
\begin{align}
\chi_{234}
& =
\frac{2^{13} \pi^3}{3^5 \, 5^3 \, 7^3}
\frac{
  \epsilon
  (\epsilon^2 - 1)
  }{
  (\epsilon^2 + 1)^4
  }
E_1^6
E_2
p_0^{24}
(\Delta p)^3
\ 
|c_\mathrm{d}(p_0)|^2
\Im\mathopen{}\Big[
  \Big\{
    4
    p_0^2
    E_1^4
    |c_\mathrm{d}(p_0)|^2
    \epsilon^2
    -
    3
    E_2^2
    |c_\mathrm{p}(p_0)|^2
    \left(\epsilon^2 + 1 \right)^2
    \nonumber \\ & \qquad \qquad \qquad \qquad \qquad 
    +
    6
    E_1^4
    c_\mathrm{s}^*(p_0)c_\mathrm{d}(p_0)
    \epsilon^2
    +
    12i
    E_1^4
    \Im(c_\mathrm{s}^*(p)c_\mathrm{d}(p))
    \left(\epsilon^2 - 1 \right)^2
    \Big\}
  \ 
  c_\mathrm{p}(p_0)c_\mathrm{d}^*(p_0)
  e^{i\varphi}
  \Big]
,
\label{PT-chi234-final-result}
\end{align}
\end{widetext}%
where the radial integration has been performed assuming that the state amplitudes are sharply peaked at $p_0$ over a small momentum interval $\Delta p$, so that they satisfy
$
\int_0^\infty
c_\mathrm{i}^*(p)
c_\mathrm{j}(p)
p^{n+k+2}
\d p
=
\Delta p
\,
c_\mathrm{i}^*(p_0)
c_\mathrm{j}(p_0)
p_0^{n+k+2}
$.

In our final result \eqref{PT-chi234-final-result} we see three distinct types of dependence of $\chi_{234}$ on the ellipticity of the fundamental,
which are proportional to
$
\displaystyle
\frac{
  \epsilon^2 - 1
  }{
  (\epsilon^2 + 1)^4
  }
\epsilon^3
$,
$
\displaystyle
\frac{
  \epsilon
  (\epsilon^2 - 1)
  }{
  (\epsilon^2 + 1)^2
  }
$,
and
$
\displaystyle
\frac{
  \epsilon
  (\epsilon^2 - 1)^3
  }{
  (\epsilon^2 + 1)^4
  }
$,
respectively,
and which combine in different proportions depending on the specific configuration used:
the radial structure of the atom and the envelope of the driving pulses determine the state constants $c_\mathrm{i}(p)$ (via~\eqref{PT-cs-cp-cd-final} and~\eqref{PT-c1-c2-def}),
and the relative amplitudes $E_1$ and $E_2$ of the driving fields then determine which of the ellipticity dependences dominate.

\section{Details of the TDSE simulations}
\label{app-tdse-details}

We perform TDSE simulations using the code from Ref.~\cite{Patchkovskii2016}. We use a radial box of size \SI{1190.34}{\au} with a log-uniform grid consisting of 3000 points. The first 10 points are on a uniform grid from \SI{0.03636}{\au} up to \SI{0.3636}{\au}, followed by 25 points on a logarithmic grid up to \SI{3.94}{\au} and 2965 points on a uniform grid until the box boundary. 
To avoid unphysical reflections, we use a complex absorber from~\cite{Manolopoulos2002} with width \SI{32.775}{\au} starting at \SI{1157.94}{\au} 
We use a timestep of \SI{0.0025}{\au} and include angular momenta up to $\ell_\mathrm{max}=30$ and all $m\in[-\ell_\mathrm{max}, \ell_\mathrm{max}]$. The photoelectron angular distributions are calculated using the iSURFC method~\cite{Morales2016}. 
The $\omega$ and $2\omega$ fields with fundamental frequency $\omega=\SI{0.375}{\au}$ (\SI{10.20}{eV}) have all the same flat-top envelopes, with 4 cycle $T=2\pi/\omega\simeq \SI{16.755}{\au}$ ($\simeq \SI{405}{as}$) long $\sin^2$ rise and fall and $8$-optical-cycle-long flat-top part. 
The peak electric field amplitude of the $\omega$ is \SI{0.005}{\au} ($\SI{8.775e11}{W/cm^2}$) while the $2\omega$ field peak amplitude is $F_{2\omega}= \SI{0.0008}{\au}$ ($\SI{2.25e10}{W/cm^2}$). The $\omega$ field has an elliptical polarization in the $xy$ plane and the $2\omega$ is linearly polarized along the ${z}$ axis.

\section{An apparent `blind spot': \texorpdfstring{$Y_{33}+Y_{43}$}{Y(33)+Y(43)}}
\label{app-blind-spot-Y33-Y43}

Coming back to more general distributions, we now turn to an apparent `blind spot' for our formalism, as mentioned briefly in the Discussion,
which is illustrative in understanding how extensions of our tensor-cross-product formalism can be used for trickier cases.
In particular, we consider a distribution of the form
\begin{align}
\rho_\mathrm{ABS}(\vbr)
& =
\Re\mathopen{}\left[
  Y_{33}(\vbr)
  +e^{i\varphi} Y_{43}(\vbr)
  \right]
f(r)
,
\label{rho-apparent-blind-spot}
\end{align}
(where $f(r)$ is a rotationally-symmetric radial amplitude) consisting of a linear superposition of octupolar and hexadecapolar terms.
This distribution, shown in Figure~\ref{fig:blind-spots}(a), forms a clearly chiral helical shape, reminiscent of the triple-gaussian helix from Section~\ref{sec-chi-234}, but without any quadrupolar moment.%
\footnote{
As an analytically-tractable alternative, \eqref{rho-apparent-blind-spot} could be replaced with combinations of the form $S_{\ell,\pm3}(\vbr)e^{-r^2}$.
}

Since we are imposing an explicitly multipolar form, only two tensorial multipolar moments are nonzero, 
namely
\begin{align}
\boldsymbol{\mu}^{(3)}_\mathrm{ABS}
=
C_3
\Re\mathopen{}\left[
  \hat{\vb{t}}_{33}
  \right]
\  \text{and} \ 
\boldsymbol{\mu}^{(4)}_\mathrm{ABS}
=
C_4
\Re\mathopen{}\left[
  e^{i\varphi}
  \hat{\vb{t}}_{43}
  \right]
,
\label{mu-3-4-for-rho-abs}
\end{align}
in terms of the multipolar basis tensors $\hat{\vb{t}}_{\ell m}$ as defined in \eqref{t-lm-basis-definition},
where 
$C_3=\sqrt{\tfrac{32\pi}{35}} \int_0^\infty f(r)r^{5}\d r$
and
$C_4=\sqrt{\tfrac{128\pi}{315}} \allowbreak{}\int_0^\infty f(r)r^{6}\d r$.
The results from section~\ref{app-tensorial-moments-poly-dist} then imply that any nonzero tensorial moments $\tens{M}^{(n)}_\mathrm{ABS}$ can only be obtained from the multipolar moments from~\eqref{mu-3-4-for-rho-abs} via the tensor lift operator $\lift$.
As such it is sufficient to consider $\boldsymbol{\mu}^{(3)}_\mathrm{ABS}$ and $\boldsymbol{\mu}^{(4)}_\mathrm{ABS}$.

As mentioned in the Discussion, the restriction to only two nonzero multipolar moment tensors then implies that we cannot use a chiral measure based directly on a triple tensor product:
while there are three nonzero pseudotensor-valued tensor cross products,
\begin{align}
\boldsymbol{\pi}^{(n)}_\mathrm{ABS}
& =
\left(
  \boldsymbol{\mu}^{(3)}_\mathrm{ABS}
  \times
  \boldsymbol{\mu}^{(4)}_\mathrm{ABS}
  \right)^{(n)}
,
\end{align}
for $n=2$, 4 and 6, those cannot be contracted with either of $\boldsymbol{\mu}^{(3)}_\mathrm{ABS}$ or $\boldsymbol{\mu}^{(4)}_\mathrm{ABS}$ to give a nonzero result.

On their own, the cross products $\boldsymbol{\pi}^{(n)}_\mathrm{ABS}$ themselves do provide a measure of asymmetry, in that their tensor norms $\boldsymbol{\pi}^{(n)}_\mathrm{ABS} \bullet \boldsymbol{\pi}^{(n)}_\mathrm{ABS}$ can only be nonzero if the distribution is chiral.
However, those norms cannot provide a sense of handedness for $\rho_\mathrm{ABS}(\vbr)$.

Fortunately, though, it \textit{is} possible to extract a sense of handedness for $\rho_\mathrm{ABS}(\vbr)$ from the cross products $\boldsymbol{\pi}^{(n)}_\mathrm{ABS}$, by referencing them against each other, 
using the \textit{even}-parity tensor triple product from~\eqref{tensor-triple-product-even-parity-def}.
This yields a pseudoscalar,
\begin{align}
\chi_{34(246)}
& =
\left(
  \boldsymbol{\pi}^{(2)}_\mathrm{ABS}
  \times
  \boldsymbol{\pi}^{(4)}_\mathrm{ABS}
  \right)^{(6)}
\tensordot
\boldsymbol{\pi}^{(6)}_\mathrm{ABS}
,
\end{align}
which can be easily computed symbolically,
\begin{align}
\chi_{34(246)}
& =
\frac{3}{1280}
C_3^3 C_4^3
\:
\sin^3(\varphi)
,
\end{align}
and which provides a clear sense of handedness for $\rho_\mathrm{ABS}(\vbr)$ based on the relative phase $\varphi$ between the octupolar and hexadecapolar components.

Structurally, this new pseudoscalar can be understood using a similar manipulation to the one used in~\eqref{chiral-moment-as-triple-integral}, substituting in the integral form of all of the $\boldsymbol{\mu}^{(\ell)}_\mathrm{ABS}$ factors.
That reveals it as a \textit{six}-point correlation function, with a pseudoscalar kernel given by an isotropic 6-point function, such as described previously for cosmological applications~\cite{Cahn2023isotropic, Varshalovich1988}.


\section{A full `blind spot': purely-octupolar chiral distributions}
\label{app-blind-spot-pure-octupole}
Finally, we turn to the `cryptochiral' distribution, mentioned in Section~\ref{sec-discussion}:
a complete `blind spot' of our formalism, and, most likely, to other existing chirality measures.
We construct this blind-spot distribution as a purely octupolar one~\cite{Gaeta2023, Chen2017}~-- that is, one sitting purely in the $\ell=3$ representation; 
for simplicity, we deal with a spherical distribution.
Generically, such a distribution has the form
\begin{align}
\rho_\mathrm{POBS}(\vbr)
& =
\sum_{m=0}^3 c_{3,m} Y_{3,m}(\vbr)
+
\mathrm{c.c.}
,
\end{align}
with complex coefficients $c_{3,m}$.
For the distribution shown in Figure~\ref{fig:blind-spots}(b), we use the coefficients
\begin{align}
(c_{3,0},c_{3,1},c_{3,2},c_{3,3})
& =
\left(
-1, \frac{i}{4}, \frac{1}{4}, 1
\right)
,
\end{align}
which can be understood as follows.

The `bones' of the distribution are provided by the tetrahedral symmetry which is normally associated with the function $\Re(Y_{3,2}(\vbr))$, which we rotate for visual clarity to align one of the main lobes with the $z$ axis;
that provides the coefficients
$
(c_{3,0},c_{3,3})
=
(-1, 1)
$.
To make this distribution chiral, we simply need to break the symmetry between the three positive lobes at the positive-$z$ side, and this will happen generically from most choices of the coefficients $c_{3,1}$ and $c_{3,2}$.
For definiteness and simplicity, we show an example with
$
(c_{3,1},c_{3,2})
=
\frac14(i,1)
$.

The chirality of the distribution as shown in Figure~\ref{fig:blind-spots}(b) is not immediately apparent, but it can be seen explicitly from the symmetry breaking of the lobes, all of which have different values of $\rho_\mathrm{POBS}(\vbr)$;
it can also be seen in the topological structure of the constant-$\rho_\mathrm{POBS}(\vbr)$ contours and how they connect the various saddle points of the distribution on the sphere.

\bibliographystyle{arthur} 
\bibliography{references}{}


\end{document}